\documentclass[longbibliography]{revtex4-1}

\usepackage{graphicx}
\usepackage{amsmath, amsfonts, amssymb, amsbsy}
\usepackage{MnSymbol}
\usepackage[usenames,dvipsnames]{xcolor}
\usepackage{marvosym}
\usepackage{sidecap}
\usepackage{ifsym}
\usepackage{color}
\usepackage{ulem}
\usepackage{soul}
\usepackage{units}

\usepackage[version=3]{mhchem}

\usepackage{siunitx,bm,float}

\usepackage{xspace}

\newcommand{\lsco} {{La$_{2-x}$Sr$_x$CuO$_4$}\@\xspace}

\newcommand{\ybco} {$\ce{YBa2Cu3O_{6+y}}$\@\xspace}

\newcommand{\tc} {\ensuremath{T_{\mathrm c}}\@\xspace}

\begin{document}
\vspace*{3cm}
\begin{center}
\textbf{\large How pressure enhances the critical temperature of superconductivity in YBa$_2$Cu$_3$O$_{6+y}$ }\\
\vspace{2cm}
{\bf \flushleft Michael Jurkutat}\\
{Felix Bloch Institute for Solid State Physics,
 Leipzig University, Linn\'{e}stra{\ss}e 5, 04103 Leipzig, Germany}\\
{\textit{now at:} Institute for Biological Interfaces 4, Karlsruhe Institute of Technology,
76344 Eggenstein-Leopoldshafen, Germany}\\
Email: m.jurkutat@gmail.com\\

{\bf \flushleft Carsten Kattinger}\\
{Felix Bloch Institute for Solid State Physics,
Leipzig University, Linn\'{e}stra{\ss}e 5, 04103 Leipzig, Germany}\\

{\bf \flushleft Stefan Tsankov}\\
{Felix Bloch Institute for Solid State Physics,
 Leipzig University, Linn\'{e}stra{\ss}e 5, 04103 Leipzig, Germany}\\

{\bf \flushleft Richard Reznicek}\\ 
{Felix Bloch Institute for Solid State Physics,
 Leipzig University, Linn\'{e}stra{\ss}e 5, 04103 Leipzig, Germany}\\
 
{\bf \flushleft Andreas Erb}\\ 
{Walther Meissner Institut, Bayerische Akademie der Wissenschaften, 85748 Garching, Germany}\\

{\bf \flushleft J\"urgen Haase}\\
{Felix Bloch Institute for Solid State Physics,
 Leipzig University, Linn\'{e}stra{\ss}e 5, 04103 Leipzig, Germany}\\
Email: j.haase@physik.uni-leipzig.de\\
Tel: +49.341.9732601\\
\end{center}
\newpage

\section*{Abstract}
High-temperature superconducting cuprates respond to doping with a dome-like dependence of their critical temperature ($T_{\mathrm{c}}$). 
But the family specific maximum $T_\mathrm{c}$ can be surpassed by application of pressure, a compelling observation known for decades. 
We investigate the phenomenon with high-pressure anvil cell nuclear magnetic resonance (NMR) and measure the charge content at planar Cu and O, and with it the doping of the ubiquitous CuO$_2$ plane with atomic scale resolution. 
We find that pressure increases the overall hole doping, as widely assumed, but when it enhances $T_\mathrm{c}$ above what can be achieved by doping, pressure leads to a hole redistribution favoring planar O. 
This is similar to the observation that the family-specific maximum $T_\mathrm{c}$ is higher for materials where the hole content at planar O is higher at the expense of that at planar Cu. 
The latter reflects a dependence of the maximum \tc on the Cu-O bond covalence and the charge-transfer gap.
The results presented here indicate that the pressure-induced enhancement of the maximum $T_\mathrm{c}$ points to the same mechanism.\\

\section*{Significance}
Understanding the maximum critical temperature of superconductivity $T_\mathrm{c,max}$ in the cuprate high-temperature superconductors is still a crucial goal, and the fact that for underdoped cuprates external pressure can lead to an enhancement of  $T_\mathrm{c,max}$ in excess of what can be achieved by doping has puzzled researchers for decades.
Here, using high-pressure nuclear magnetic resonance (NMR) we measure the local charges at planar Cu and O, and we find that pressure indeed changes hole-doping, but, in addition, it can lead to a redistribution of holes between O and Cu. The latter effect is in line with previous NMR analyses at ambient pressure, that revealed that $T_\mathrm{c,max}$ in different cuprate material families is nearly proportional to the O hole content, which is understood to reflect the crucial role of the charge transfer gap and the Cu-O bond covalence. 
We show that the enigmatic \tc enhancement by pressure points to the same origin.

\section{Introduction}
High-temperature superconducting cuprates \cite{Bednorz1986} are still central to condensed matter physics, and they carry surprisingly rich electronic properties \cite{Keimer2015} despite sharing a rather simple CuO$_2$ plane as a common structural unit. 
By doping the antiferromagnetic parent materials with electrons or holes these properties are induced, in particular superconductivity with its critical temperature ($T_\mathrm{c}$) that shows a dome-like dependence on the doping level. 
However, while the maximum value $T_\mathrm{c,max}$ appears at the so-called optimal doping levels (near $\sim 16\%$) for all materials, the family-dependent $T_\mathrm{c,max}$ differ widely. 

\begin{figure}[t]
  \begin{center}
    \includegraphics[width=0.48\textwidth]{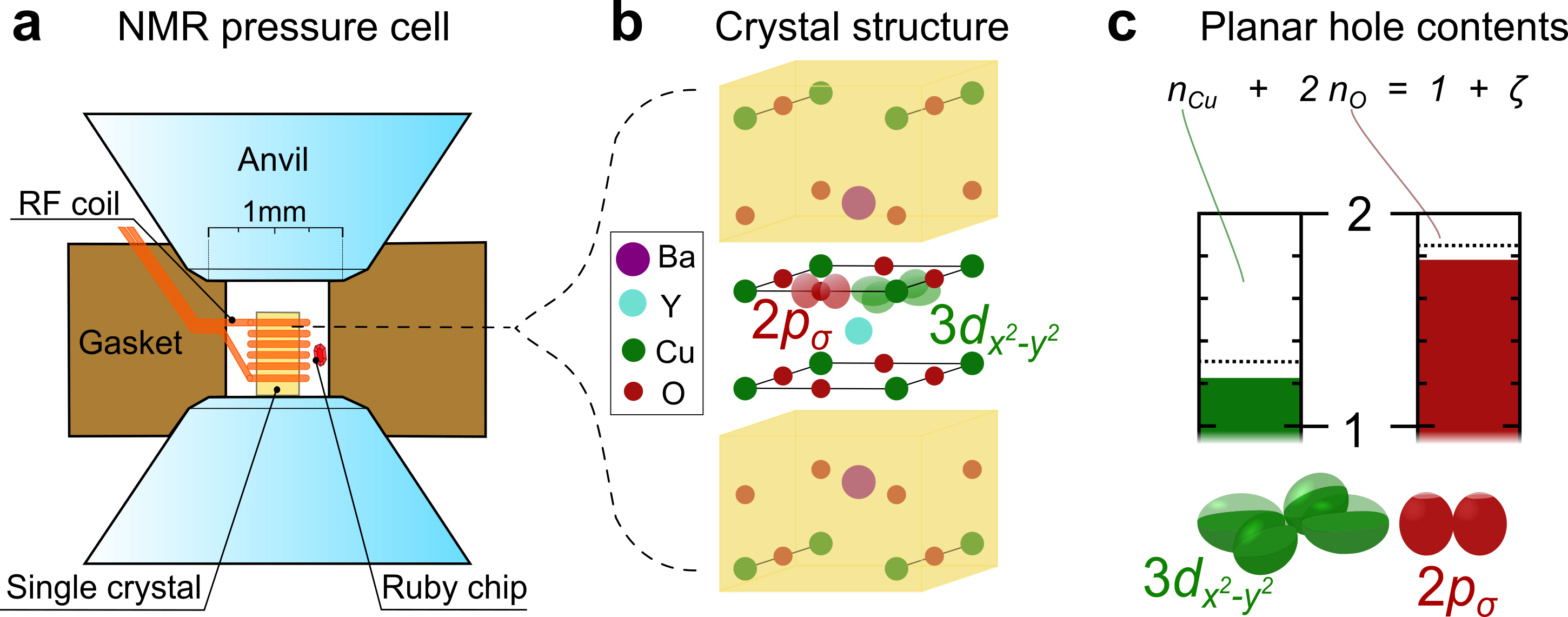}
    \caption{{\bf Anvil cell high-pressure excerted on YBa$_2$Cu$_3$O$_{6+y}$ changes the charges in the CuO$_2$ plane}. \textbf{(a)} Schematic of the anvil cell used for NMR; the micro-coil surrounds the single crystal of about 1 nano-L volume, and both are placed in the high pressure chamber with a ruby chip as an optical pressure gauge. 
    \textbf{(b)} Sketch of the crystal structure of YBa$_2$Cu$_3$O$_{6+y}$ with highlighted bonding orbitals in one of the CuO$_2$ planes. \textbf{(c)} The hole content of these bonding orbitals can be measured with Cu and O NMR quadrupole splittings, see \textit{Methods}. From the measured hole contents for Cu ($n_{\mathrm{Cu}}$) and O ($n_{\mathrm{O}}$) the total doping measured NMR, $\zeta$, follows ($1 +  \zeta = n_{\mathrm{Cu}} + 2 n_{\mathrm{O}}$).}
    \label{fig:f1}
  \end{center}
\end{figure}

\begin{figure*}[t]
\includegraphics[width=.9\columnwidth]{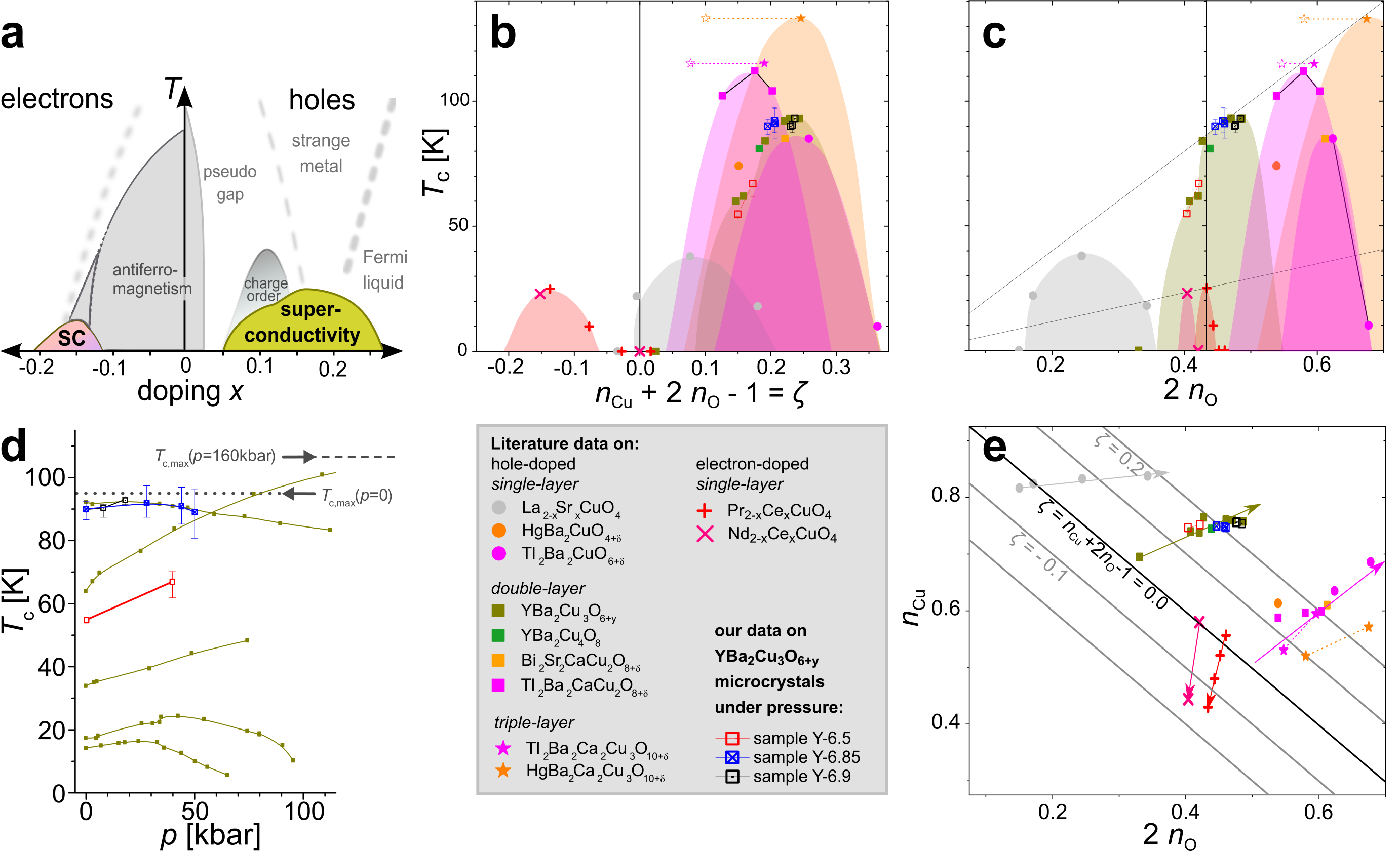}
\caption{\label{fig:f2} 
\textbf{Charges in the CuO$_2$ plane and $T_\mathrm{c}$} 
\textbf{(a)} An electronic phase diagram is typically used to mark the various cuprate phenomena as a function of doping ($x$) and temperature ($T$), however, important details like the maximum \tc differ between material families and are not set by $x$. 
\textbf{(b)} $T_\mathrm{c}$ as a function of $\zeta=n_{\mathrm{Cu}}+2n_{\mathrm{O}}-1$, i.e., the doping measured with NMR. 
\textbf{(c)} $T_\mathrm{c}$ as a function of $2n_{\mathrm{O}}$ orders the superconducting domes; a near proportionality between $T_\mathrm{c,max}$ and $n_\mathrm{O}$ is revealed \cite{Rybicki2016}. 
\textbf{(e)} The planar charge distribution in terms of $n_{\mathrm{Cu}}$ and $2n_{\mathrm{O}}$ \cite{Jurkutat2014} reveals significant differences between the various families. 
This display relates to the above panels, sharing the abscissa with \textbf{c} and indicating diagonal lines of constant doping $\zeta$ corresponding to the abscissa of \textbf{b}.
\textbf{(d)} $T_\mathrm{c}$ vs. pressure for different doping levels of YBa$_2$Cu$_3$O$_{6+y}$ (YBCO) for the samples measure here and literature data, cf. legend. $T_\mathrm{c}$ slowly decreases for optimally doped YBCO with pressure, while $T_\mathrm{c}$ increases for underdoped YBCO, and can even exceed the maximum $T_\mathrm{c}$ achievable with chemical doping (gray lines). Results from this work are shown for three different materials together with literature data in panels \textbf{b}-\textbf{e}, cf. legend.
}
\end{figure*}

Previously, it was shown that nuclear magnetic resonance (NMR) can measure the charges in the CuO$_2$ plane at the atomic level, i.e., in terms of the Cu ($n_\mathrm{Cu}$) and O ($n_\mathrm{O}$) bonding orbital hole contents, and a simple relation was found \cite{Jurkutat2014},
\begin{equation}\label{eq:1}
1 + \zeta = n_\mathrm{Cu} + 2 n_\mathrm{O}.
\end{equation}
This relation is expected if $\zeta$ is similar to the chemical doping that adds to the hole already present in the parent compound, where Cu is nominally in 3d$^9$ configuration. 
Interestingly, it was found that the actual sharing of the hole content between Cu and O appears to be a fundamental parameter for \tc: $T_\mathrm{c,max}$ is nearly proportional to $n_\mathrm{O}$ \cite{Rybicki2016}, which is largely determined by the parent chemistry. 
So to increase $T_\mathrm{c,max}$ electron charge has to be transferred from planar O to planar Cu, an experimental correlation that holds for all known hole-doped cuprates, $T_\mathrm{c,max} \approx \SI{200}{K}\cdot 2n_\mathrm{O}$ \cite{Rybicki2016}, cf. also Fig.~\ref{fig:f1}b-c and Fig.~\ref{fig:f2}a-c,e.
This identifies the material chemistry parameters controlling the hole sharing betweeen Cu and O - the charge transfer gap and bond covalence - as crucial in determining the maximumum \tc obtainable at optimal doping.
An observation that only recently could be reproduced theoretically by solving the three-band Hubbard model.\cite{Kowalski2021}

This intriguing relation between $T_\mathrm{c,max}$ and the hole sharing in the CuO$_2$ plane may hold information about another mystery of cuprate behavior:\@ the unusual, material-specific \textit{pressure} dependence of $T_\mathrm{c}$ \cite{Lorenz2005,Schilling2007}. 
While the pressure response of $T_\mathrm{c}$ is different for the various cuprate families, quite generally, $T_\mathrm{c}$ of underdoped cuprates tends to increase with pressure ($p$), while it hardly changes for optimal doping and usually decreases for overdoped materials \cite{Schilling1992,Sadewasser2000}.  
This suggests that pressure increases planar hole doping, which is also supported by conductivity and Hall measurements \cite{Schilling1992,Lorenz2005,Murayama1991,Schilling2007}. 
However, some underdoped samples show a significantly higher $T_\mathrm{c,max}(p)$ than what can be achieved by chemical doping ($x$), as is the case for YBa$_2$Cu$_3$O$_{6+y}$ investigated here, cf. Fig.~\ref{fig:f2}\,d.
This phenomenon is of great interest as it may relate to the mechanism of superconductivity.

Clearly, with the NMR results mentioned above, it appears intriguing to explore how pressure affects the charges in the CuO$_2$ plane at the atomic level. 
However, since this requires $^{63}$Cu and $^{17}$O NMR experiments on oriented single crystals at pressures that can only be achieved with anvil cell devices, such experiments are rather challenging: a micro-crystal surrounded by a radio frequency micro-coil needs to be positioned inside the pressurized region of an anvil cell, as depicted in Fig.~\ref{fig:f1}a. 
Apart from mechanical issues such as disruptive failures induced by a changing geometry with pressure, for example, the signal-to-noise for the aligning process of the micro-crystal with respect to the magnetic field is a limiting factor. 
We were able to meet these requirements to some extent and report on the results obtained for a few micro-crystals of YBa$_2$Cu$_3$O$_{6+y}$ at pressures up to about \SI{44}{kbar}. 
We find that pressure indeed has two effects: it increases the overall hole doping in the plane, as is generally expected, but it also changes the sharing of the holes between Cu and O, and can thereby increase \tc as well. 
These findings underline the importance of the sharing of the planar charges for \tc, and the therein reflected role of the charge transfer gap and the Cu-O bond covalence.\cite{Kowalski2021}\par\medskip

\begin{figure*}[]
\includegraphics[width=.7\columnwidth]{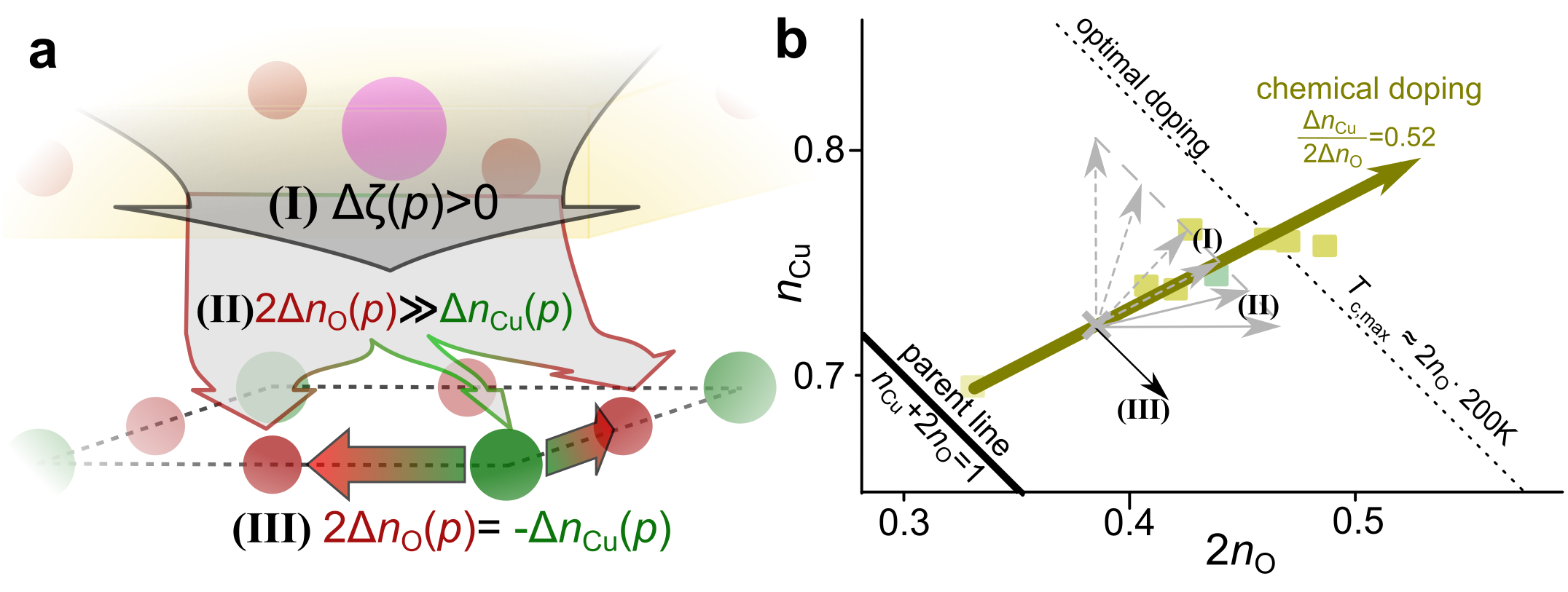}
\caption{\label{fig:f3} \textbf{Pressure effects on planar charges}. \textbf{(a)} Pressure can induce hole doping of the CuO$_2$ plane (I), and the holes arrive predominantly at planar O (II), but pressure can also induce intra-planar hole redistribution from Cu to O (III). 
\textbf{(b)} The same effects are described in the 'YBCO-region' of the $n_\mathrm{Cu}$ - $2n_\mathrm{O}$ plane from Fig.~\ref{fig:f2}e, with the parent line indicated in black. 
An underdoped ($ x \sim 10\%$) YBCO system would be located near ($n_\mathrm{Cu}$,$2n_\mathrm{O}$)=(0.72,0.38), indicated by a grey cross. 
If pressure (I) increases hole doping to a certain level (dashed line parallel to the optimal doping line), the system could follow any of the grey arrows. If (II) pressure favors hole doping of planar O more than chemical doping, a shallower slope is expected (full gray arrows). (III) If the charges redistributed within the plane, the system would follow the black full arrow given the overall doping remained the same.}
\end{figure*}

\section{Planar Charge Distribution and \tc under pressure}
The common electronic phase diagram of the cuprates, cf. Fig.~\ref{fig:f2}a, assumes that chemical doping ($x$) is the key variable.
Since we can measure the NMR doping level $\zeta$ given by \eqref{eq:1}, we prefer to use $\zeta$ as the actual doping level, if NMR measurements are available.
To keep the discussion transparent we use the variable $x$ if chemical doping is known from other sources, e.g. by using the superconducting dome or from stoichiometry. 
Slight differences between the two numbers ($\zeta$ and $x$) become apparent by noting that the superconducting domes do not fall exactly on top of each other in a $T$-$\zeta$ phase diagram, cf.~Fig.~\ref{fig:f2}b.

The sharing of charges between Cu and O in the CuO$_2$ plane (at ambient conditions) is reproduced in Fig.~\ref{fig:f2}e \cite{Jurkutat2014}, where the black diagonal 'parent' line ($\zeta = 0$) separates the hole-doped regime above ($\zeta > 0$) from the electron-doped below it ($\zeta < 0$). 
The various cuprate families then start at very different points near the parent line. 
These starting points also determine the ratio ($\Delta n_\mathrm{Cu}/2\Delta n_\mathrm{O}$) of how the doped holes entering the plane are distributed (slopes indicated by arrows). 
This is because sharing of both, inherent and doped charges are determined by the material-dependent planar bond covalence.

We now focus on YBa$_2$Cu$_3$O$_{6+y}$ (YBCO), full dark yellow squares in Fig.~\ref{fig:f2}e or in more detail in Fig.~\ref{fig:f3}b. 
In the undoped ($y=0$) material YBa$_{2}$Cu$_3$O$_6$, the inherent hole must be shared  between Cu and O, and we estimate $n_\mathrm{Cu}=0.68$ and $2n_\mathrm{O}=0.32$ from this plot.
Upon doping, the holes enter the CuO$_2$ plane as indicated by the dark yellow arrow that points away from the parent line with the slope of $\Delta n_\mathrm{Cu}/2\Delta n_\mathrm{O}\approx 0.52$, cf.~Fig.~\ref{fig:f3}b. 
The question this work aims to address is how pressure affects the planar charges $n_\mathrm{Cu}$ and $n_\mathrm{O}$.

In Fig.~\ref{fig:f3} we show schematically what could be expected if pressure enhances \tc by increasing the O hole content.
Given the general pressure dependence of \tc as well as conductivity and Hall measurements \cite{Schilling1992,Lorenz2005,Murayama1991,Schilling2007}, one expects that (I) pressure increases planar hole content, $\Delta_p \zeta > 0$.
(II) To increase the maximum \tc by boosting the O hole content under pressure, clearly pressure-induced hole-doping should favor O more than chemical doping does, i.e. $\Delta_p n_\mathrm{Cu}/2\Delta_p n_\mathrm{O} < 0.52$.
This would require a decrease in charge transfer gap and an increase in bond covalence, which would (III) cause an intra-planar hole-transfer from Cu to O, although this effect could be masked by coincident pressure-induced doping. 

In the literature \cite{Neumeier1993} the change of $T_\mathrm{c}$ with pressure is typically described phenomenologically as:
\begin{equation}\label{eq:2}
\left( \frac{d T_c }{d p} \right)_\mathrm{tot} =\left( \frac{\partial T_c }{\partial x} \right) \left( \frac{\partial x }{\partial p} \right) + \left( \frac{d T_c }{d p} \right)_\mathrm{intr}
\end{equation}
The  first term on the r.h.s. describes the change of doping due to pressure $(\partial x/ \partial p)$ with $\partial T_\mathrm{c}/ \partial x$ given by the slope of the superconducting dome as a function of doping at ambient pressure. 
The second term $( dT_\mathrm{c} / dp)_\mathrm{intr}$ describes the (unknown) intrinsic pressure effects on $T_\mathrm{c}$, i.e., pressure-induced change of the shape of the superconducting dome. 
Although \eqref{eq:2} is not necessary for our analysis, we will discuss our results also in this context.

\section{High pressure NMR Experiments}
In order to measure the planar charges under pressure, high-pressure $^{63}$Cu and $^{17}$O anvil cell NMR experiments were performed with home-made anvil cells \cite{Kattinger2021} that fit standard NMR magnets (\SI{11.7}{T} and \SI{17.6}{T}) and home-made probes.
Therefore, our anvil cells are rather small compared to what is used by another group \cite{Kitagawa2010} that also engages in single crystal NMR experiments (of other materials) at similar pressures. 
We use $^{17}$O exchanged small volume (0.3 to  1.5 nano-L) micro-crystals with 3 different stoichiometries: YBa$_2$Cu$_3$O$_{6.5}$ (Y-6.5),  YBa$_2$Cu$_3$O$_{6.85}$ (Y-6.85), and YBa$_2$Cu$_3$O$_{6.9}$ (Y-6.9). 
These doping levels were originally determined from \tc measurements (see Supplementary Sec.~3).  
The crystals were glued on one of the anvil's culets, and RF micro-coils were placed around them with the leads fed to outside the pressurized region through channels carved in the gasket, paraffin oil ensured hydrostatic conditions. 
Pressure was applied with a hydraulic press and screws secured the pressure during NMR experiments.

Standard orientation dependent NMR experiments were performed to measure the quadrupole frequencies (splittings of the Zeeman resonance) for $^{63}$Cu and $^{17}$O in the CuO$_2$ plane, from which the hole densities can be determined (see Methods). 
The NMR doping levels at ambient pressure for the samples used here are $\zeta = 0.15, 0.19, 0.23$ for Y-6.5, Y-6.85, and Y-6.9, respectively.

\begin{figure}
\includegraphics[width=0.5\columnwidth]{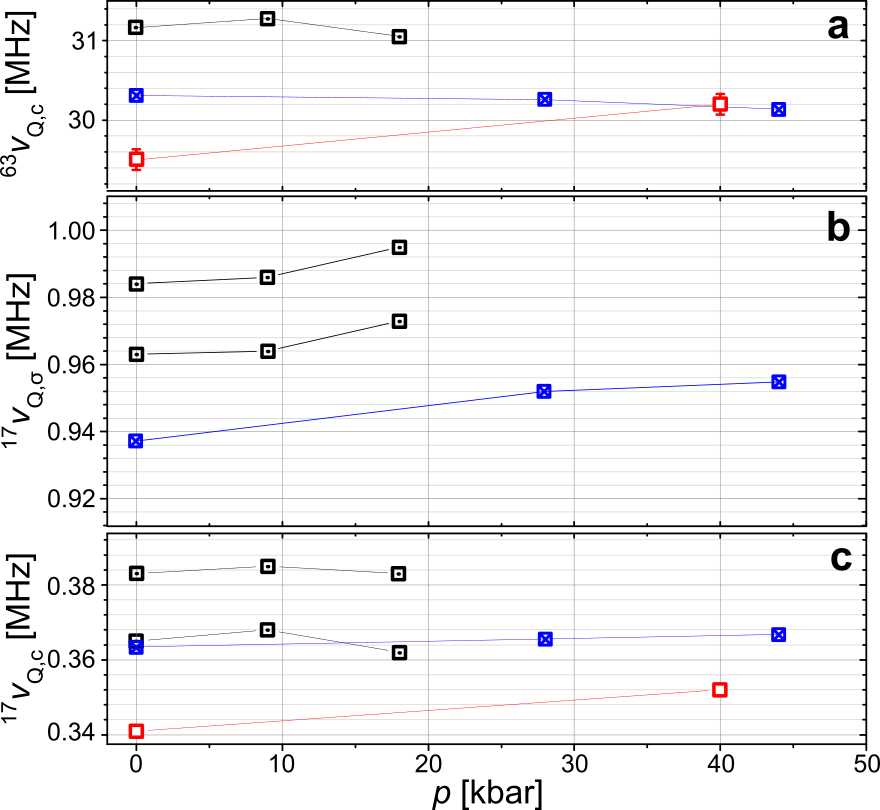}
\caption{\label{fig:f4} \textbf{Experimental data}. Pressure dependence of the $^{17}$O and $^{63}$Cu quadrupole frequencies (${^{17,63}\nu_{Q,\alpha}}$) for the aligned single crystals of Y-6.9 (dotted black squares), Y-6.85 (crossed blue squares) and Y-6.5 (red open squares); $\alpha$ denotes the direction of the external magnetic field $B_0$. \textbf{(a)} $^{63}$Cu along the crystal $c$-axis, \textbf{(b)} $^{17}$O along the $\sigma$-bond direction, and \textbf{(c)} $^{17}$O along the crystal $c$-axis. For Y-6.9 both quadrupole frequencies reflecting the double peak feature of the satellite transitions are displayed (the error, indicated, is typically much less than the symbol size).}
\end{figure}

The measured pressure dependence of the NMR quadrupole frequencies (${^{63,17}\nu}_\mathrm{Q}$) of the aligned single crystals are summarized in Fig.~\ref{fig:f4}.
For planar Cu in Fig.~\ref{fig:f4}a we find that $^{63}\nu_\mathrm{Q,c}$ increases for the underdoped Y-6.5, but it is less sensitive to pressure in the higher doped Y-6.85 and Y-6.9, and even slightly decreases at elevated pressure.
Both observations are consistent with previous Cu NMR reports on underdoped and optimally doped YBCO \cite{Brinkmann1992}.

For planar O in Fig.~\ref{fig:f4}b-c we find that $^{17}\nu_\mathrm{Q,c}$ (field along the crystal $c$-direction) and $^{17}\nu_\mathrm{Q,\sigma}$ (field along the Cu-O $\sigma$-bond) generally increase with pressure for all doping levels, although this is more pronounced for the underdoped Y-6.5.
While the literature on $^{17}$O NMR in cuprates under pressure is limited, one study on single crystals of underdoped YBCO up to \SI{18}{kbar} found increasing $^{17}$O quadrupole splittings as well \cite{Vinograd2019}.
We note that the peculiar changes in splittings for Y-6.9 have been shown to signify charge ordering in that compound at elevated pressure.\cite{Reichardt2018}

Using the pressure-induced changes of the $^{63}$Cu and $^{17}$O NMR quadrupole splittings depicted in Fig.~\ref{fig:f4}, we determined the planar charges as a function of pressure, see \textit{Methods} \eqref{e:17O} and \eqref{e:63Cu}.

\section{Planar charges under pressure}

\begin{figure*}[]
\includegraphics[width=.9\columnwidth]{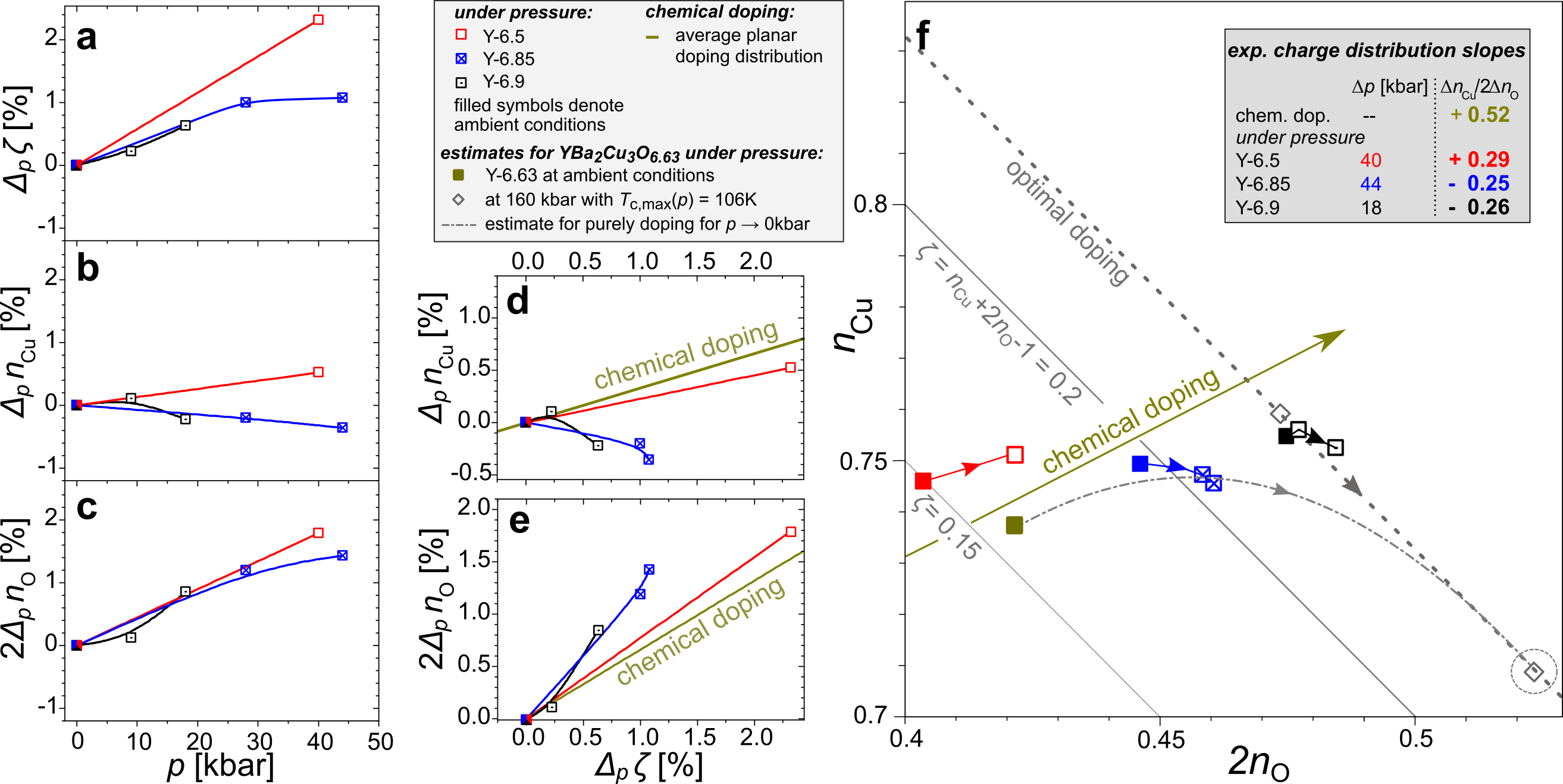}
\caption{\label{fig:f5} \textbf{Planar charge distribution as determined by NMR} for samples Y-6.9 (black), Y-6.85 (blue) and Y-6.5 (red). 
\textbf{(a)} All samples show increasing hole-doping $\Delta_p \zeta = \Delta_p n_\mathrm{Cu} + 2\Delta_p n_\mathrm{O} > 0$ with pressure. 
The slope $\Delta_p \zeta / \Delta p$ is higher in underdoped Y-6.5. 
\textbf{(b)} The Cu hole content is found to increase for the underdoped sample Y-6.5, $\Delta_p n_\mathrm{Cu}>0$ , this is weaker for higher doping and $n_\mathrm{Cu}$ even decreases at elevated pressure. 
\textbf{(c)} The O hole content $2n_\mathrm{O}$ increases for all samples similarly, with approximately $\SI{4 e-4}{holes/kbar}$. 
In order to  compare pressure effects to chemical doping (dark yellow lines), we show $\Delta_p n_\mathrm{Cu}$ and $2 \Delta_p n_\mathrm{O} $ as a function of pressure-induced doping $\Delta_p \zeta$ in \textbf{(d)} and \textbf{(e)}, respectively. 
The increase in Cu hole content, $\Delta_p n_\mathrm{Cu}$, is smaller for pressure-induced doping compared with that induced by chemical doping for all samples. 
The O hole content $2\Delta_p n_\mathrm{O}$ increases much faster with pressure-induced doping compared to chemical doping for all samples. 
For both Cu and O, the underdoped Y-6.5 is closest to chemical doping, where the higher doped Y-6.9 and Y-6.85 also show an intra-planar charge redistribution, i.e. an increase of O holes at the expense of Cu holes. 
\textbf{(f)} Zoom into the ($n_\mathrm{Cu}, 2n_\mathrm{O}$)-plane (cf.~Figs.~\ref{fig:f2}e and \ref{fig:f3}b) near our experimental data (full symbols denote ambient pressure data, and full lines with arrows indicate increasing pressure). 
Also shown are literature data for YBa$_2$Cu$_3$O$_{6.63}$ and the estimated high-pressure point (\SI{160}{kbar}, $T_\mathrm{c} = \SI{106}{K}$, circled empty diamond) is that of an optimally doped YBCO with appropriate charge redistribution for the enhanced $T_\mathrm{c}$ ($\Delta T_\mathrm{c,max} = \SI{11}{K}, 2\Delta n_\mathrm{O} \approx 5.5\%$). 
The dash-dotted gray line is a parabolic fit between the two points, $\Delta_p n_\mathrm{Cu} = 0.52 \cdot 2\Delta_p n_\mathrm{O} - 8 \cdot (2\Delta_p n_\mathrm{O})^2$, where the linear slope 0.52 is defined by the chemical doping.}
\end{figure*}

The pressure-induced changes ($\Delta_p$) in the average local hole contents ($\Delta_p n_{\mathrm{Cu}}$ and $\Delta_p n_{\mathrm{O}}$) add up to the total change in hole content ($\Delta_p \zeta$) of the CuO$_2$ plane, cf. \eqref{eq:1}.
We find $\Delta_p \zeta > 0$ for all samples, i.e., we observe an increase in hole doping with increasing pressure, cf. Fig.~\ref{fig:f5}a. 
This hole doping is more pronounced for underdoped Y-6.5 with an initial slope of $\approx\SI{5.8 e-4}{holes/kbar}$, compared to only $\approx\SI{3.5 e-4}{holes/kbar}$ for near optimally doped Y-6.85 as well as Y-6.9. 

However, the changes of the site-specific hole contents with pressure differ between materials, as can be seen in Fig.~\ref{fig:f5}b-c. 
While $\Delta_p n_\mathrm{Cu}/p \SI{\approx 1.3 e-4}{holes/kbar}$ for underdoped Y-6.5, the materials closer to optimal doping, Y-6.85 and Y-6.9, show a much weaker or no increase at lower pressure and even a decrease in Cu hole content beyond \SI{10}{kbar}, cf. Fig.~\ref{fig:f5}b.
For O we find that pressure causes a similar increase for all three samples, i.e.\@ $2 \Delta_p n_\mathrm{O}/p \approx \SI{4 e-4}{holes/kbar}$, cf. Fig.~\ref{fig:f5}c.
This clearly indicates a pressure-induced intra-planar charge redistribution, certainly for the higher doped Y-6.85 and Y-6.9, where the O hole content increases stronger than doping, i.e., $2 \Delta_p n_\mathrm{O}  > \Delta_p \zeta $, and the Cu hole content decreases $\Delta_p n_\mathrm{Cu} < 0$.

In order to compare pressure effects to chemical doping, the pressure-induced changes of the Cu and O hole contents as a function of pressure-induced doping ($\Delta_p \zeta$) are shown in Fig.~\ref{fig:f5}d-e. 
We observe a relative decrease of $n_\mathrm{Cu}$ and an increase of  $n_\mathrm{O}$, compared to what is found for chemically doped charges.

To summarize, our high-pressure NMR experiments on YBCO have shown that the increase in $T_\mathrm{c}$ with pressure is accompanied by changes in the local hole contents that lead to an increased $n_\mathrm{O}$ compared to chemical doping.
All three effects displayed in Fig.~\ref{fig:f3} were observed: (I) an increase in hole doping, $\Delta_p \zeta > 0$, that (II) favors an increase in O holes ($n_\mathrm{O}$) over those at Cu ($n_\mathrm{Cu}$).
And at high doping levels and elevated pressure even (III) intraplanar charge redistribution can be observed.

Clearly, the data qualitatively indicate that pressure induces hole doping as well as an increase in bond covalence.
Before discussing our results more broadly, we first consider whether the observed changes in planar charges are quantitatively sufficient to account for the pressure-enhanced maximum \tc reported for YBCO as shown in Fig.~\ref{fig:f2}d. 
Unfortunately, YBa$_2$Cu$_3$O$_{6.63}$, which shows the highest pressure-induced increase in $T_\mathrm{c}$, is not part of our final set of samples. 
In addition, we are lacking data for much higher pressures with our single crystal anvil NMR. 
However, from our samples with doping levels below and above that of YBa$_2$Cu$_3$O$_{6.63}$ we can nonetheless obtain a quantitative estimate. 
The literature data for YBa$_2$Cu$_3$O$_{6.63}$ show that  $T_\mathrm{c}$ increases from about \SI{64}{K} at ambient pressure to about \SI{106}{K} at \SI{160}{kbar}.
This means an increase of $T_\mathrm{c,max}$ of about \SI{11}{K} compared to that of the optimally doped material. 
According to the experimental relation, $T_\mathrm{c,max} \approx \SI{200}{K}\cdot 2n_\mathrm{O}$, this requires an increase of $2n_\mathrm{O}$ by 5.5\% for an optimally doped YBCO. 
We can find the position of such a material in the ($n_\mathrm{Cu}, 2n_\mathrm{O}$)-plot by following a line of constant doping, beginning at chemically optimally doped YBCO, until we reach the encircled, empty diamond in the lower right corner of Fig.~\ref{fig:f5}f. 
Applying pressure means that the hole contents move from ($n_\mathrm{Cu}, 2n_\mathrm{O}$) = (0.738, 0.423) to (0.709, 0.524), cf.~Fig.~\ref{fig:f5}f. 
So under pressure of \SI{160}{kbar} the O hole content in YBa$_2$Cu$_3$O$_{6.63}$ has to increase by 10\%, i.e. at a rate of $2 \Delta_p n_\mathrm{O}/p \approx \SI{6.3 e-4}{holes/kbar}$, which is comparable to the average increase in O hole content we see for our samples of $\SI{4 e-4}{holes/kbar}$. 
While we do not have data on YBa$_2$Cu$_3$O$_{6.63}$ and only reach one fourth of the necessary pressure to unlock its  $T_\mathrm{c}(p)$ peak, we do show a possible path for illustration purposes.
To reproduce the chemical doping-like hole distribution for lower doping and lower pressure, we assume a parabolic dependence that leads to the empty diamond and has an initial slope given by the chemical doping. 
We obtain the dash-dotted line in Fig.~\ref{fig:f5}f, which reproduces the overall features of our experimental data quite well.

\section{Discussion}

Pressure-induced doping clearly depends on the material and chemical doping, and previous assessments range widely with maximum values up to \SI{0.2}{\%/kbar} \cite{Wijngaarden1999,Ambrosch-Draxl2004}, while we find that $\partial \zeta / \partial p \approx \SI{0.058(5)}{\%/kbar}$ for the underdoped Y-6.5, and \SI{0.036(5)}{\%/kbar} for the samples near optimal doping. 
A recent estimate by Alireza et al.\@ \cite{Alireza2017} of \SI{0.032}{\%/kbar} for fully doped YBCO matches our results quite well.
Note that our data imply a pressure-induced doping, $\partial \zeta / \partial p $, that is stronger for underdoped YBCO, contrary to modelling assumptions used elsewhere \cite{Cyr-Choi2018,Vinograd2019}.

Pressure favors a higher O hole content $2 n_\mathrm{O}$ compared to what can be achieved by chemical doping to the extent that, particularly at higher pressure and for higher doping levels, $2 n_\mathrm{O}$ increases not only through doping but also at the expense of a decreasing Cu hole content ($n_\mathrm{Cu}$). 
A similar effect was recently reported with first principle calculations for Bi-based cuprates by Deng et al. \cite{Deng2019}. 
Their results for a pressure-induced increase of Cu 3$d(x^2-y^2)$ occupation and so a decrease in Cu hole content of $\partial n_\mathrm{Cu} / \partial p =-0.04\%$/kbar is  more pronounced than what we find, cf. Fig.~\ref{fig:f5}b.
The pressure-induced decrease in $n_\mathrm{Cu}$, while simultaneously overall doping increases, clearly reveals an intra-planar charge redistribution under pressure.

The sharing of the inherent hole that is nominally on Cu and the distribution of additional (chemically) doped charges, both reveal Cu and O contributions to occupied and unoccupied electronic states.
Depending on the context in which cuprates are discussed, this reflects the Cu-O bond covalence, the charge transfer gap, or Cu and O band contributions.
An intra-planar redistribution of holes from Cu to O therefore signals an increase in Cu-O bond covalence, i.e., a decrease in the charge transfer gap and an increased contribution of O to unoccupied bands and of Cu to occupied bands.
The concurrent increase in $T_\mathrm{c,max}$ under pressure is consistent with the proportionality between $T_\mathrm{c,max}$ and the O hole content seen by NMR.
Studies using other methods also suggest an increasing $T_\mathrm{c,max}$ with a decreasing charge transfer gap \cite{Weber2012,Ruan2016,Weber2017}.
Recently, Kowalski et al.\cite{Kowalski2021} solved the three-band Hubbard model with parameters that capture the variable charge transfer gap and bond covalence. 
Their results reproduce the varying, material-dependent O hole contents found with NMR \cite{Jurkutat2014} that scale with the maximum \tc\cite{Rybicki2016}. 
Kowalski et al. also find that the optimal doping level increases with decreasing charge transfer gap and increasing $T_\mathrm{c,max}$, which, interestingly, fits the trend of mismatching domes in Fig.~\ref{fig:f2}b. 

Our results suggest that the sought-after \textit{intrinsic} effect of pressure on $T_\mathrm{c}$, cf.~\eqref{eq:2}, is a decrease of the charge transfer gap, i.e. an increase in planar Cu-O bond covalence.
Although our sample set and pressure range were limited, a simple estimate for the necessary changes of the planar charge contents under pressure in underdoped YBa$_2$Cu$_3$O$_{6.63}$ shows quantitative agreement with the changes in planar charges we find.
Also Sadewasser et al. \cite{Sadewasser1999} estimated for the intrinsic pressure effect on $T_\mathrm{c}$ in YBCO about \SI{0.1}{K/kbar}.
The data here show an increase of $2n_\mathrm{O}$ under pressure for all samples of about \SI{0.042(6)}{\%/kbar}, cf. Fig.~\ref{fig:f5}c. 
When multiplied with the slope of the $T_\mathrm{c,max}/(2 n_\mathrm{O}) \approx \SI{200}{K/hole}$, this gives \SI{0.084(1)}{K/kbar}, in good agreement with \cite{Sadewasser1999}.

While our results qualitatively and quantitatively account for the intrinsic pressure effect that increases $T_\mathrm{c,max}$ in YBCO, the pressure phenomenology of $T_\mathrm{c}$ differs somewhat for different cuprate families.
Clearly, the specific crystal structure and doping level should have an influence on how much pressure affects doping and changes planar bonding.

For \lsco{}, for instance, $T_\mathrm{c}$ increases with pressure for all doping levels, indicating that pressure causes an intra-planar charge redistribution that increases (decreases) planar O (Cu) hole content and has hardly any effect on doping. 
The latter is also consistent with the pressure-independent Hall coefficient for all doping levels of this family \cite{Murayama1991}.

For the Bi-, Tl- and Hg-based cuprate materials that can be realized in single-layer as well as different multi-layer configurations, the pressure phenomenology is much more complex, e.g., including non-monotonic $T_\mathrm{c}$-dependence on pressure for some materials.
However, an interesting question concerns triple-layer materials (and beyond), as these exhibit distinct outer and inner CuO$_2$ layers and, under pressure, can exhibit two maxima in $T_\mathrm{c}$.
Perhaps, this relates to different effects of pressure on the different layers in terms of intra-planar and inter-planar charge distribution as well as doping. 
The latter effect has already been indicated by first-principle calculations \cite{Ambrosch-Draxl2004}.

The weak $T_\mathrm{c}$-dependence on pressure in optimally electron-doped materials \cite{Murayama1989,Markert1990,Ishiwata2013} could be accounted for by similar effects as in YBCO, i.e., compensating effects on $T_\mathrm{c}$ with pressure increasing $T_\mathrm{c,max}$ while also pushing the system to the underdoped regime through hole doping. 

Finally, we would like to emphasize that both the previously reported proportionality between $T_\mathrm{c,max}$ and planar O hole content for different cuprate families \cite{Jurkutat2014,Rybicki2016}, and the increase of $T_\mathrm{c,max}$ under pressure by increasing planar O hole content reported here, do not give any explanation for the peak of $T_\mathrm{c}$ at optimal doping and the superconducting dome. 
Only the height of the latter, $T_\mathrm{c,max}$, as well as other cuprate properties \cite{Jurkutat2019b} appear to be fundamentally linked to the role of O in the planar structure.

The role of O holes and bond covalence has to be of crucial importance for material chemistry as well as any theoretical attempt at understanding cuprate superconductivity.

\section{Methods}

\subsection{Sample preparation}
High-quality single crystals of \ybco were grown in non-reactive BaZrO$_3$ crucibles and annealed as described elsewhere \cite{Erb1996}. The resulting fully oxygenated single crystals (y = 1) were twinned within the a-b plane. 
For the Y-6.9 sample, a micro-crystal of an approximate size of $150 \times 100 \times 100 \mu m^3$ was cut from the slab and subsequently $^{17}$O exchanged, as previously described \cite{Reichardt2018}, which results in nearly optimally doped YBCO. 
In order to produce the $^{17}$O enriched underdoped samples Y-6.5 and Y-6.85, we exchanged larger single crystals with $^{17}$O and subsequently annealed them to obtain the desired chain O content. They were cut into micro-crystals afterwards.

Prior to inserting the crystals into the pressure cell, the crystal axes were determined by polarized light that can easily identify domain boundaries in the twinned $a$-$b$-plane at the surface. The crystals were fixed to one of the culet surfaces with epoxy so that the $c$-axis is nearly parallel to the culet surface [see Supplementary, Fig. 4a].
After closing the pressure cell, the $T_\mathrm{c}$ of the enclosed sample was determined using a NMR probe with a cryostat in zero field. The circuit was tuned at about \SI{200}{MHz} at a temperature slightly above $T_\mathrm{c}$. Then, the temperature was lowered throughout the superconducting transition and the concomitant change of the tank circuit frequency was monitored; the process was repeated by starting below $T_\mathrm{c}$ and raising the temperature. $T_\mathrm{c}$ was defined as the upper temperature where about 10\% of the rapid frequency shift had occurred [see Supplementary, Fig. 5].\\

\subsection{Pressure cell preparation}
Our home-built pressure cells have cylindrical cell bodies with a diameter of about $\SI{17}{mm}$ and a height of about  $\SI{20}{mm}$ [see Supplementary, Fig. 1a]. The cell body is made from titanium. 
Optical access to the sample region is possible due to transparent anvils (along the cell axis) and 3 drilled holes in the cell body in radial direction at angles of $120^{\circ}$. The latter allow for an inspection of the anvils and the gasket while the cell is closed to avoid destruction of the single crystal. The ruby luminescence technique was used to measure the pressure through the axial hole \cite{Forman1972}.
Further details on the preparation of the cell, including the gasket, can be found elsewhere \cite{Meier2014}.

\subsection{NMR experiments}
For the experiments commercial Bruker or Tecmag pulse spectrometers were used with $\SI{11.7}{T}$ or $\SI{17.6}{T}$ superconducting magnets. The anvil cells were mounted on regular home-made probes that fit commercial cryostats for temperature variation. Spin echo ($\pi/2$-$\tau$-$\pi$) pulse sequences were employed, and if possible, whole transitions were excited and recorded, while frequency stepped echoes were employed for broad lines. 
The $\pi/2$ pulse length for a typical experiment was accordingly $\SI{0.5}{\mu s}$ or $\SI{7}{\mu s}$. The average pulse power varied between $\SI{10}{mW}$ and $\SI{5}{W}$ (note that the small volume of the RF micro-coils requires rather low power levels).

Different RF micro-coil designs were tested, with various filling and Q factors, according to different sizes and shapes of the crystals. 
The micro-coils were wound from an insulated silver wire (Goodfellow Cambridge Ltd) with diameter of $\SI{25}{\mu m}$ ($\SI{5}{\mu m}$ insulation). 
The DC resistances measured on the closed cells were found to vary between $\sim\SI{0.7}{\Omega} \text{ and } \sim\SI{1.5}{\Omega}$ at room temperature (the lead resistances are smaller due to a larger diameter). 
With a typical coil inductance of $\SI{50}{nH}$, this is in agreement with the measured Q factors that ranged between $20$ and $40$ (the RF skin depth is similar to the radius of the wire).

For the first cell (Y-6.5 crystal) we used micro-coil with nearly elliptical cross-section to increase the filling factor. 
The crystal itself was extremely flat and small. 
It had the dimensions of approximately $90 \times 90 \times  \SI{40}{\mu m^3}$. 
The filling factor was about 0.13 [see Supplementary Fig. 3b]. 

For the second cell (Y-6.85 crystal), a double-wound micro-coil was used with a higher inductance and greater mechanical stability [see Supplementary, Fig. 3a]. 
The dimension of the crystal was $ 140 \times 140 \times \SI{90}{\mu m^3}$. 
The filling factor of this coil was about 0.3. 

For the third cell (Y-6.9 crystal) a regular cylindrical coil was used.
The crystal had the dimensions $ 150 \times 100 \times  \SI{100}{\mu m^3}$. 
The filling factor of the coil was estimated to be about 0.4. \par\medskip

Since the signal-to-noise ratio (SNR) is critical, the noise was always measured and verified that it is of thermal origin, predominantly from the RF micro-coil (an overall noise figure of about $\SI{1.25}{dB}$ was determined at room temperature).

The highest SNR (per scan) measured (in the time domain) on the central transition of planar $^{63}$Cu for ${c \parallel B_0}$ at room temperature and a bandwidth of 5 MHz was $SNR = \SI{4.9e-2}{}$ for the Y-6.9 cell; for the Y-6.85 and Y-6.5 cell the SNR was about $\SI{3.2e-2}{}$ and $\SI{0.4e-2}{}$, respectively. 
For the planar O central transition, at a bandwidth of 2 MHz, we found SNRs of $\SI{2.9e-2}{}$,  $\SI{1.8e-2}{} $, and $\SI{0.12e-2}{}$ for Y-6.9, Y-6.85, and Y-6.5, respectively.
With the necessary repetition times, a single spectrum could require 24 hours of signal averaging.
Due to the low signal (and SNR) for Y-6.5, only a limited set of data was recorded.
Nutation experiments were performed to find the pulse lengths that were close (within factor of two) to the estimated RF amplitudes.

For the orientation of a cell with respect to the magnetic field $B_0$ a goniometer that was mounted on the home-built NMR probe was used [see Supplementary, Fig.~1b, Fig.~2].
While the single crystals were glued to one anvil with the $c$-axis parallel to its culet surface, the true crystal orientation was measured with the goniometer that holds the anvil cell \cite{Kattinger2021}. 
If the satellite linewidths and SNRs permitted, the satellite resonances were followed as a function of angles, cf. \cite{Reichardt2018}. Otherwise, angular dependences for the planar Cu central transition were recorded.

The full angular dependence of the Cu NMR central transition of the Y-6.5 cell is shown in the Supplementary [see Supplementary Fig. 4b].

\subsection*{Determination of charges}
The Cu and O splittings along the respective principle axes are related to the planar hole densities as follows \cite{Haase2004,Jurkutat2014}:
\begin{equation}
^{17}\nu_\mathrm{Q,\sigma} = 2.45 \mathrm{MHz} \cdot n_\mathrm{O} + 0.39 \mathrm{MHz}
\label{e:17O}
\end{equation}
\begin{equation}
^{63}\nu_\mathrm{Q,c} = 94.3 \mathrm{MHz} \cdot n_\mathrm{Cu} - 5.68 \mathrm{MHz} \cdot (8 - 4 n_\mathrm{O})
\label{e:63Cu}
\end{equation}

In the case of the Y-6.5 sample only splittings in $c$-direction could be measured, where the changes of the splitting are only half of what is observed along the bond, i.e., $\Delta_p ^{17}\nu_\mathrm{Q,c} = 2.45 \mathrm{MHz} /2 \cdot \Delta_p n_\mathrm{O}$. 

In order to determine $n_\mathrm{O}$ from $^{17}\nu_\mathrm{Q,\sigma}$ for the initial chemical doping level for this sample we took literature data summarized in \cite{Haase2004} on $^{17}\nu_\mathrm{Q,c}$, $^{17}\nu_\mathrm{Q,\sigma}$, $T_\mathrm{c}$ and O content for various doping levels of YBa$_2$Cu$_3$O$_{6+y}$.

\section{Author contribution}
M.J. led NMR experiments and data analysis, supported by C.K., S.T., J.H. Materials were supplied by A.E. who supervised also sample preparation. C.K. led the anvil cell construction with help from R.R. and J.H. All authors contributed to writing of the manuscript, which was led by M.J., C.K. and J.H. The overall project leadership was with J.H.

\begin{acknowledgments}
 
\textbf{Acknowledgement} We acknowledge the financial support by the Deutsche Forschungsgemeinschaft, Project No. 317319632, and by Leipzig University.\\
 
\end{acknowledgments}

\bibliography{MichaPressure_pnas}

\begin{thebibliography}{33}%
\makeatletter
\providecommand \@ifxundefined [1]{%
 \@ifx{#1\undefined}
}%
\providecommand \@ifnum [1]{%
 \ifnum #1\expandafter \@firstoftwo
 \else \expandafter \@secondoftwo
 \fi
}%
\providecommand \@ifx [1]{%
 \ifx #1\expandafter \@firstoftwo
 \else \expandafter \@secondoftwo
 \fi
}%
\providecommand \natexlab [1]{#1}%
\providecommand \enquote  [1]{``#1''}%
\providecommand \bibnamefont  [1]{#1}%
\providecommand \bibfnamefont [1]{#1}%
\providecommand \citenamefont [1]{#1}%
\providecommand \href@noop [0]{\@secondoftwo}%
\providecommand \href [0]{\begingroup \@sanitize@url \@href}%
\providecommand \@href[1]{\@@startlink{#1}\@@href}%
\providecommand \@@href[1]{\endgroup#1\@@endlink}%
\providecommand \@sanitize@url [0]{\catcode `\\12\catcode `\$12\catcode
  `\&12\catcode `\#12\catcode `\^12\catcode `\_12\catcode `\%12\relax}%
\providecommand \@@startlink[1]{}%
\providecommand \@@endlink[0]{}%
\providecommand \url  [0]{\begingroup\@sanitize@url \@url }%
\providecommand \@url [1]{\endgroup\@href {#1}{\urlprefix }}%
\providecommand \urlprefix  [0]{URL }%
\providecommand \Eprint [0]{\href }%
\providecommand \doibase [0]{http://dx.doi.org/}%
\providecommand \selectlanguage [0]{\@gobble}%
\providecommand \bibinfo  [0]{\@secondoftwo}%
\providecommand \bibfield  [0]{\@secondoftwo}%
\providecommand \translation [1]{[#1]}%
\providecommand \BibitemOpen [0]{}%
\providecommand \bibitemStop [0]{}%
\providecommand \bibitemNoStop [0]{.\EOS\space}%
\providecommand \EOS [0]{\spacefactor3000\relax}%
\providecommand \BibitemShut  [1]{\csname bibitem#1\endcsname}%
\let\auto@bib@innerbib\@empty
\bibitem [{\citenamefont {Bednorz}\ and\ \citenamefont
  {M{\"{u}}ller}(1986)}]{Bednorz1986}%
  \BibitemOpen
  \bibfield  {author} {\bibinfo {author} {\bibfnamefont {J.~G.}\ \bibnamefont
  {Bednorz}}\ and\ \bibinfo {author} {\bibfnamefont {K.~A.}\ \bibnamefont
  {M{\"{u}}ller}},\ }\bibfield  {title} {\enquote {\bibinfo {title} {{Possible
  high $T_\text{c}$ superconductivity in the Ba-La-Cu-O system}},}\ }\href@noop
  {} {\bibfield  {journal} {\bibinfo  {journal} {Z. Phys. B Condens. Matter}\
  }\textbf {\bibinfo {volume} {193}},\ \bibinfo {pages} {189--193} (\bibinfo
  {year} {1986})}\BibitemShut {NoStop}%
\bibitem [{\citenamefont {Keimer}\ \emph {et~al.}(2015)\citenamefont {Keimer},
  \citenamefont {Kivelson}, \citenamefont {Norman}, \citenamefont {Uchida},\
  and\ \citenamefont {Zaanen}}]{Keimer2015}%
  \BibitemOpen
  \bibfield  {author} {\bibinfo {author} {\bibfnamefont {B.}~\bibnamefont
  {Keimer}}, \bibinfo {author} {\bibfnamefont {S.~A.}\ \bibnamefont
  {Kivelson}}, \bibinfo {author} {\bibfnamefont {M.~R.}\ \bibnamefont
  {Norman}}, \bibinfo {author} {\bibfnamefont {S.}~\bibnamefont {Uchida}}, \
  and\ \bibinfo {author} {\bibfnamefont {J.}~\bibnamefont {Zaanen}},\
  }\bibfield  {title} {\enquote {\bibinfo {title} {{From quantum matter to
  high-temperature superconductivity in copper oxides}},}\ }\href {\doibase
  10.1038/nature14165} {\bibfield  {journal} {\bibinfo  {journal} {Nature}\
  }\textbf {\bibinfo {volume} {518}},\ \bibinfo {pages} {179--186} (\bibinfo
  {year} {2015})}\BibitemShut {NoStop}%
\bibitem [{\citenamefont {Jurkutat}\ \emph {et~al.}(2014)\citenamefont
  {Jurkutat}, \citenamefont {Rybicki}, \citenamefont {Sushkov}, \citenamefont
  {Williams}, \citenamefont {Erb},\ and\ \citenamefont {Haase}}]{Jurkutat2014}%
  \BibitemOpen
  \bibfield  {author} {\bibinfo {author} {\bibfnamefont {Michael}\ \bibnamefont
  {Jurkutat}}, \bibinfo {author} {\bibfnamefont {Damian}\ \bibnamefont
  {Rybicki}}, \bibinfo {author} {\bibfnamefont {Oleg~P}\ \bibnamefont
  {Sushkov}}, \bibinfo {author} {\bibfnamefont {Grant V~M}\ \bibnamefont
  {Williams}}, \bibinfo {author} {\bibfnamefont {Andreas}\ \bibnamefont {Erb}},
  \ and\ \bibinfo {author} {\bibfnamefont {J{\"{u}}rgen}\ \bibnamefont
  {Haase}},\ }\bibfield  {title} {\enquote {\bibinfo {title} {{Distribution of
  electrons and holes in cuprate superconductors as determined from $^{17}$O
  and $^{63}$Cu nuclear magnetic resonance}},}\ }\href@noop {} {\bibfield
  {journal} {\bibinfo  {journal} {Phys. Rev. B}\ }\textbf {\bibinfo {volume}
  {90}},\ \bibinfo {pages} {140504} (\bibinfo {year} {2014})}\BibitemShut
  {NoStop}%
\bibitem [{\citenamefont {Rybicki}\ \emph {et~al.}(2016)\citenamefont
  {Rybicki}, \citenamefont {Jurkutat}, \citenamefont {Reichardt}, \citenamefont
  {Kapusta},\ and\ \citenamefont {Haase}}]{Rybicki2016}%
  \BibitemOpen
  \bibfield  {author} {\bibinfo {author} {\bibfnamefont {Damian}\ \bibnamefont
  {Rybicki}}, \bibinfo {author} {\bibfnamefont {Michael}\ \bibnamefont
  {Jurkutat}}, \bibinfo {author} {\bibfnamefont {Steven}\ \bibnamefont
  {Reichardt}}, \bibinfo {author} {\bibfnamefont {Czeslaw}\ \bibnamefont
  {Kapusta}}, \ and\ \bibinfo {author} {\bibfnamefont {Juergen}\ \bibnamefont
  {Haase}},\ }\bibfield  {title} {\enquote {\bibinfo {title} {{Perspective on
  the phase diagram of cuprate high-temperature superconductors}},}\
  }\href@noop {} {\bibfield  {journal} {\bibinfo  {journal} {Nat. Commun.}\
  }\textbf {\bibinfo {volume} {7}},\ \bibinfo {pages} {1--6} (\bibinfo {year}
  {2016})}\BibitemShut {NoStop}%
\bibitem [{\citenamefont {Kowalski}\ \emph {et~al.}(2021)\citenamefont
  {Kowalski}, \citenamefont {Dash}, \citenamefont {S\'emon}, \citenamefont
  {S\'en\'echal},\ and\ \citenamefont {Tremblay}}]{Kowalski2021}%
  \BibitemOpen
  \bibfield  {author} {\bibinfo {author} {\bibfnamefont {Nicolas}\ \bibnamefont
  {Kowalski}}, \bibinfo {author} {\bibfnamefont {Sidhartha~Shankar}\
  \bibnamefont {Dash}}, \bibinfo {author} {\bibfnamefont {Patrick}\
  \bibnamefont {S\'emon}}, \bibinfo {author} {\bibfnamefont {David}\
  \bibnamefont {S\'en\'echal}}, \ and\ \bibinfo {author} {\bibfnamefont
  {Andr\'e-Marie}\ \bibnamefont {Tremblay}},\ }\bibfield  {title} {\enquote
  {\bibinfo {title} {Oxygen hole content, charge-transfer gap, covalency, and
  cuprate superconductivity},}\ }\href@noop {} {\bibfield  {journal} {\bibinfo
  {journal} {Proceedings of the National Academy of Sciences}\ }\textbf
  {\bibinfo {volume} {118}},\ \bibinfo {pages} {e2106476118} (\bibinfo {year}
  {2021})}\BibitemShut {NoStop}%
\bibitem [{\citenamefont {Lorenz}\ and\ \citenamefont
  {Chu}(2005)}]{Lorenz2005}%
  \BibitemOpen
  \bibfield  {author} {\bibinfo {author} {\bibfnamefont {B.}~\bibnamefont
  {Lorenz}}\ and\ \bibinfo {author} {\bibfnamefont {C.W.}\ \bibnamefont
  {Chu}},\ }\bibfield  {title} {\enquote {\bibinfo {title} {{High Pressure
  Effects on Superconductivity}},}\ }in\ \href@noop {} {\emph {\bibinfo
  {booktitle} {Frontiers in Superconducting Materials}}}\ (\bibinfo
  {publisher} {Springer-Verlag},\ \bibinfo {address} {Berlin/Heidelberg},\
  \bibinfo {year} {2005})\ pp.\ \bibinfo {pages} {459--497}\BibitemShut
  {NoStop}%
\bibitem [{\citenamefont {Schilling}(2007)}]{Schilling2007}%
  \BibitemOpen
  \bibfield  {author} {\bibinfo {author} {\bibfnamefont {J~S}\ \bibnamefont
  {Schilling}},\ }\bibfield  {title} {\enquote {\bibinfo {title}
  {{High-Pressure Effects}},}\ }in\ \href@noop {} {\emph {\bibinfo {booktitle}
  {Handbook of High-Temperature Superconductivity}}},\ \bibinfo {editor}
  {edited by\ \bibinfo {editor} {\bibfnamefont {J~R}\ \bibnamefont
  {Schrieffer}}}\ (\bibinfo  {publisher} {Springer},\ \bibinfo {year}
  {2007})\BibitemShut {NoStop}%
\bibitem [{\citenamefont {Schilling}\ and\ \citenamefont
  {Klotz}(1992)}]{Schilling1992}%
  \BibitemOpen
  \bibfield  {author} {\bibinfo {author} {\bibfnamefont {James~S.}\
  \bibnamefont {Schilling}}\ and\ \bibinfo {author} {\bibfnamefont {Stefan}\
  \bibnamefont {Klotz}},\ }\bibfield  {title} {\enquote {\bibinfo {title} {{The
  Influence of High Pressure on the Superconducting and Normal Properties of
  High Temperature Superconductors}},}\ }in\ \href@noop {} {\emph {\bibinfo
  {booktitle} {Physical Properties of High Temperature Superconductors III}}}\
  (\bibinfo  {publisher} {World Scientific},\ \bibinfo {year} {1992})\ pp.\
  \bibinfo {pages} {59--157}\BibitemShut {NoStop}%
\bibitem [{\citenamefont {Sadewasser}\ \emph {et~al.}(2000)\citenamefont
  {Sadewasser}, \citenamefont {Schilling}, \citenamefont {Paulikas},\ and\
  \citenamefont {Veal}}]{Sadewasser2000}%
  \BibitemOpen
  \bibfield  {author} {\bibinfo {author} {\bibfnamefont {S}~\bibnamefont
  {Sadewasser}}, \bibinfo {author} {\bibfnamefont {J~S}\ \bibnamefont
  {Schilling}}, \bibinfo {author} {\bibfnamefont {A~P}\ \bibnamefont
  {Paulikas}}, \ and\ \bibinfo {author} {\bibfnamefont {B~W}\ \bibnamefont
  {Veal}},\ }\bibfield  {title} {\enquote {\bibinfo {title} {{Pressure
  dependence of $T_\text{c}$ to 17 GPa with and without relaxation effects in
  superconducting YBa2Cu3O\textit{x}}},}\ }\href@noop {} {\bibfield  {journal}
  {\bibinfo  {journal} {Phys. Rev. B}\ }\textbf {\bibinfo {volume} {61}},\
  \bibinfo {pages} {741} (\bibinfo {year} {2000})}\BibitemShut {NoStop}%
\bibitem [{\citenamefont {Murayama}\ \emph {et~al.}(1991)\citenamefont
  {Murayama}, \citenamefont {Iye}, \citenamefont {Enomoto}, \citenamefont
  {M{\^{o}}ri}, \citenamefont {Yamada}, \citenamefont {Matsumoto},
  \citenamefont {Kubo}, \citenamefont {Shimakawa},\ and\ \citenamefont
  {Manako}}]{Murayama1991}%
  \BibitemOpen
  \bibfield  {author} {\bibinfo {author} {\bibfnamefont {C.}~\bibnamefont
  {Murayama}}, \bibinfo {author} {\bibfnamefont {Y.}~\bibnamefont {Iye}},
  \bibinfo {author} {\bibfnamefont {T.}~\bibnamefont {Enomoto}}, \bibinfo
  {author} {\bibfnamefont {N.}~\bibnamefont {M{\^{o}}ri}}, \bibinfo {author}
  {\bibfnamefont {Y.}~\bibnamefont {Yamada}}, \bibinfo {author} {\bibfnamefont
  {T.}~\bibnamefont {Matsumoto}}, \bibinfo {author} {\bibfnamefont
  {Y.}~\bibnamefont {Kubo}}, \bibinfo {author} {\bibfnamefont {Y.}~\bibnamefont
  {Shimakawa}}, \ and\ \bibinfo {author} {\bibfnamefont {T.}~\bibnamefont
  {Manako}},\ }\bibfield  {title} {\enquote {\bibinfo {title} {{Correlation
  between the pressure-induced changes in the Hall coefficient and $T_\text{c}$
  in superconducting cuprates}},}\ }\href@noop {} {\bibfield  {journal}
  {\bibinfo  {journal} {Physica C: Superconductivity}\ }\textbf {\bibinfo
  {volume} {183}},\ \bibinfo {pages} {277--285} (\bibinfo {year}
  {1991})}\BibitemShut {NoStop}%
\bibitem [{\citenamefont {Neumeier}\ and\ \citenamefont
  {Zimmermann}(1993)}]{Neumeier1993}%
  \BibitemOpen
  \bibfield  {author} {\bibinfo {author} {\bibfnamefont {J~J}\ \bibnamefont
  {Neumeier}}\ and\ \bibinfo {author} {\bibfnamefont {H~A}\ \bibnamefont
  {Zimmermann}},\ }\bibfield  {title} {\enquote {\bibinfo {title} {{Pressure
  dependence of the superconducting transition temperature of YBa2Cu3O7 as a
  function of carrier concentration: A test for a simple charge-transfer
  model}},}\ }\href@noop {} {\bibfield  {journal} {\bibinfo  {journal} {Phys.
  Rev. B}\ }\textbf {\bibinfo {volume} {47}},\ \bibinfo {pages} {8385--8388}
  (\bibinfo {year} {1993})}\BibitemShut {NoStop}%
\bibitem [{\citenamefont {Kattinger}\ \emph {et~al.}(2021)\citenamefont
  {Kattinger}, \citenamefont {Guehne}, \citenamefont {Tsankov}, \citenamefont
  {Jurkutat}, \citenamefont {Erb},\ and\ \citenamefont
  {Haase}}]{Kattinger2021}%
  \BibitemOpen
  \bibfield  {author} {\bibinfo {author} {\bibfnamefont {Carsten}\ \bibnamefont
  {Kattinger}}, \bibinfo {author} {\bibfnamefont {Robin}\ \bibnamefont
  {Guehne}}, \bibinfo {author} {\bibfnamefont {Stefan}\ \bibnamefont
  {Tsankov}}, \bibinfo {author} {\bibfnamefont {Michael}\ \bibnamefont
  {Jurkutat}}, \bibinfo {author} {\bibfnamefont {Andreas}\ \bibnamefont {Erb}},
  \ and\ \bibinfo {author} {\bibfnamefont {Juergen}\ \bibnamefont {Haase}},\
  }\bibfield  {title} {\enquote {\bibinfo {title} {High-pressure single crystal
  {NMR} in anvil cells},}\ }\href@noop {} {\bibfield  {journal} {\bibinfo
  {journal} {Rev. Sci. Instrum.}\ }\textbf {\bibinfo {volume} {92}},\ \bibinfo
  {pages} {113901} (\bibinfo {year} {2021})}\BibitemShut {NoStop}%
\bibitem [{\citenamefont {Kitagawa}\ \emph {et~al.}(2010)\citenamefont
  {Kitagawa}, \citenamefont {Gotou}, \citenamefont {Yagi}, \citenamefont
  {Yamada}, \citenamefont {Matsumoto}, \citenamefont {Uwatoko},\ and\
  \citenamefont {Takigawa}}]{Kitagawa2010}%
  \BibitemOpen
  \bibfield  {author} {\bibinfo {author} {\bibfnamefont {Kentaro}\ \bibnamefont
  {Kitagawa}}, \bibinfo {author} {\bibfnamefont {Hirotada}\ \bibnamefont
  {Gotou}}, \bibinfo {author} {\bibfnamefont {Takehiko}\ \bibnamefont {Yagi}},
  \bibinfo {author} {\bibfnamefont {Atsushi}\ \bibnamefont {Yamada}}, \bibinfo
  {author} {\bibfnamefont {Takehiko}\ \bibnamefont {Matsumoto}}, \bibinfo
  {author} {\bibfnamefont {Yoshiya}\ \bibnamefont {Uwatoko}}, \ and\ \bibinfo
  {author} {\bibfnamefont {Masashi}\ \bibnamefont {Takigawa}},\ }\bibfield
  {title} {\enquote {\bibinfo {title} {Space efficient opposed-anvil
  high-pressure cell and its application to optical and {NMR} measurements up
  to 9 {GPa}},}\ }\href@noop {} {\bibfield  {journal} {\bibinfo  {journal} {J.
  Phys. Soc. Jap.}\ }\textbf {\bibinfo {volume} {79}},\ \bibinfo {pages}
  {024001} (\bibinfo {year} {2010})}\BibitemShut {NoStop}%
\bibitem [{\citenamefont {Brinkmann}(1992)}]{Brinkmann1992}%
  \BibitemOpen
  \bibfield  {author} {\bibinfo {author} {\bibfnamefont {D}~\bibnamefont
  {Brinkmann}},\ }\bibfield  {title} {\enquote {\bibinfo {title} {{Comparing
  Y-Ba-Cu-O superconductors by Cu, O and Ba NMR/NQR}},}\ }\href@noop {}
  {\bibfield  {journal} {\bibinfo  {journal} {Applied Magnetic Resonance}\
  }\textbf {\bibinfo {volume} {3}},\ \bibinfo {pages} {483--494} (\bibinfo
  {year} {1992})}\BibitemShut {NoStop}%
\bibitem [{\citenamefont {Vinograd}\ \emph {et~al.}(2019)\citenamefont
  {Vinograd}, \citenamefont {Zhou}, \citenamefont {Mayaffre}, \citenamefont
  {Kr\"amer}, \citenamefont {Liang}, \citenamefont {Hardy}, \citenamefont
  {Bonn},\ and\ \citenamefont {Julien}}]{Vinograd2019}%
  \BibitemOpen
  \bibfield  {author} {\bibinfo {author} {\bibfnamefont {I.}~\bibnamefont
  {Vinograd}}, \bibinfo {author} {\bibfnamefont {R.}~\bibnamefont {Zhou}},
  \bibinfo {author} {\bibfnamefont {H.}~\bibnamefont {Mayaffre}}, \bibinfo
  {author} {\bibfnamefont {S.}~\bibnamefont {Kr\"amer}}, \bibinfo {author}
  {\bibfnamefont {R.}~\bibnamefont {Liang}}, \bibinfo {author} {\bibfnamefont
  {W.~N.}\ \bibnamefont {Hardy}}, \bibinfo {author} {\bibfnamefont {D.~A.}\
  \bibnamefont {Bonn}}, \ and\ \bibinfo {author} {\bibfnamefont {M.-H.}\
  \bibnamefont {Julien}},\ }\bibfield  {title} {\enquote {\bibinfo {title}
  {{Nuclear magnetic resonance study of charge density waves under hydrostatic
  pressure in YBa2Cu3O\textit{y}}},}\ }\href@noop {} {\bibfield  {journal}
  {\bibinfo  {journal} {Phys. Rev. B}\ }\textbf {\bibinfo {volume} {100}},\
  \bibinfo {pages} {094502} (\bibinfo {year} {2019})}\BibitemShut {NoStop}%
\bibitem [{\citenamefont {Reichardt}\ \emph {et~al.}(2018)\citenamefont
  {Reichardt}, \citenamefont {Jurkutat}, \citenamefont {Guehne}, \citenamefont
  {Kohlrautz}, \citenamefont {Erb},\ and\ \citenamefont
  {Haase}}]{Reichardt2018}%
  \BibitemOpen
  \bibfield  {author} {\bibinfo {author} {\bibfnamefont {Steven}\ \bibnamefont
  {Reichardt}}, \bibinfo {author} {\bibfnamefont {Michael}\ \bibnamefont
  {Jurkutat}}, \bibinfo {author} {\bibfnamefont {Robin}\ \bibnamefont
  {Guehne}}, \bibinfo {author} {\bibfnamefont {Jonas}\ \bibnamefont
  {Kohlrautz}}, \bibinfo {author} {\bibfnamefont {Andreas}\ \bibnamefont
  {Erb}}, \ and\ \bibinfo {author} {\bibfnamefont {J{\"u}rgen}\ \bibnamefont
  {Haase}},\ }\bibfield  {title} {\enquote {\bibinfo {title} {{Bulk charge
  ordering in the CuO2 plane of the cuprate superconductor YBa2Cu3O6.9 by
  high-pressure NMR}},}\ }\href@noop {} {\bibfield  {journal} {\bibinfo
  {journal} {Condens. Matter}\ }\textbf {\bibinfo {volume} {3}},\ \bibinfo
  {pages} {23} (\bibinfo {year} {2018})}\BibitemShut {NoStop}%
\bibitem [{\citenamefont {Wijngaarden}\ \emph {et~al.}(1999)\citenamefont
  {Wijngaarden}, \citenamefont {{Tristan Jover}},\ and\ \citenamefont
  {Griessen}}]{Wijngaarden1999}%
  \BibitemOpen
  \bibfield  {author} {\bibinfo {author} {\bibfnamefont {Rinke~J}\ \bibnamefont
  {Wijngaarden}}, \bibinfo {author} {\bibfnamefont {D}~\bibnamefont {{Tristan
  Jover}}}, \ and\ \bibinfo {author} {\bibfnamefont {R}~\bibnamefont
  {Griessen}},\ }\bibfield  {title} {\enquote {\bibinfo {title} {{Intrinsic and
  carrier density effects on the pressure dependence of $T_\text{c}$ of
  high-temperature superconductors}},}\ }\href@noop {} {\bibfield  {journal}
  {\bibinfo  {journal} {Physica B: Condensed Matter}\ }\textbf {\bibinfo
  {volume} {265}},\ \bibinfo {pages} {128 -- 135} (\bibinfo {year}
  {1999})}\BibitemShut {NoStop}%
\bibitem [{\citenamefont {Ambrosch-Draxl}\ \emph {et~al.}(2004)\citenamefont
  {Ambrosch-Draxl}, \citenamefont {Sherman}, \citenamefont {Auer},\ and\
  \citenamefont {Thonhauser}}]{Ambrosch-Draxl2004}%
  \BibitemOpen
  \bibfield  {author} {\bibinfo {author} {\bibfnamefont {C.}~\bibnamefont
  {Ambrosch-Draxl}}, \bibinfo {author} {\bibfnamefont {E.~Ya.}\ \bibnamefont
  {Sherman}}, \bibinfo {author} {\bibfnamefont {H.}~\bibnamefont {Auer}}, \
  and\ \bibinfo {author} {\bibfnamefont {T.}~\bibnamefont {Thonhauser}},\
  }\bibfield  {title} {\enquote {\bibinfo {title} {{Pressure-induced hole
  doping of the Hg-based cuprate superconductors}},}\ }\href@noop {} {\bibfield
   {journal} {\bibinfo  {journal} {Phys. Rev. Lett.}\ }\textbf {\bibinfo
  {volume} {92}},\ \bibinfo {pages} {187004} (\bibinfo {year}
  {2004})}\BibitemShut {NoStop}%
\bibitem [{\citenamefont {Alireza}\ \emph {et~al.}(2017)\citenamefont
  {Alireza}, \citenamefont {Zhang}, \citenamefont {Guo}, \citenamefont
  {Porras}, \citenamefont {Loew}, \citenamefont {Hsu}, \citenamefont
  {Lonzarich}, \citenamefont {Le~Tacon}, \citenamefont {Keimer},\ and\
  \citenamefont {Sebastian}}]{Alireza2017}%
  \BibitemOpen
  \bibfield  {author} {\bibinfo {author} {\bibfnamefont {P.~L.}\ \bibnamefont
  {Alireza}}, \bibinfo {author} {\bibfnamefont {G.~H.}\ \bibnamefont {Zhang}},
  \bibinfo {author} {\bibfnamefont {W.}~\bibnamefont {Guo}}, \bibinfo {author}
  {\bibfnamefont {J.}~\bibnamefont {Porras}}, \bibinfo {author} {\bibfnamefont
  {T.}~\bibnamefont {Loew}}, \bibinfo {author} {\bibfnamefont {Y.-T.}\
  \bibnamefont {Hsu}}, \bibinfo {author} {\bibfnamefont {G.~G.}\ \bibnamefont
  {Lonzarich}}, \bibinfo {author} {\bibfnamefont {M.}~\bibnamefont {Le~Tacon}},
  \bibinfo {author} {\bibfnamefont {B.}~\bibnamefont {Keimer}}, \ and\ \bibinfo
  {author} {\bibfnamefont {Suchitra~E.}\ \bibnamefont {Sebastian}},\ }\bibfield
   {title} {\enquote {\bibinfo {title} {{Accessing the entire overdoped regime
  in pristine ${\mathrm{YBa}}_{2}{\mathrm{Cu}}_{3}{\mathrm{O}}_{6+x}$ by
  application of pressure}},}\ }\href@noop {} {\bibfield  {journal} {\bibinfo
  {journal} {Phys. Rev. B}\ }\textbf {\bibinfo {volume} {95}},\ \bibinfo
  {pages} {100505} (\bibinfo {year} {2017})}\BibitemShut {NoStop}%
\bibitem [{\citenamefont {Cyr-Choini\`ere}\ \emph {et~al.}(2018)\citenamefont
  {Cyr-Choini\`ere}, \citenamefont {LeBoeuf}, \citenamefont {Badoux},
  \citenamefont {Dufour-Beaus\'ejour}, \citenamefont {Bonn}, \citenamefont
  {Hardy}, \citenamefont {Liang}, \citenamefont {Graf}, \citenamefont
  {Doiron-Leyraud},\ and\ \citenamefont {Taillefer}}]{Cyr-Choi2018}%
  \BibitemOpen
  \bibfield  {author} {\bibinfo {author} {\bibfnamefont {O.}~\bibnamefont
  {Cyr-Choini\`ere}}, \bibinfo {author} {\bibfnamefont {D.}~\bibnamefont
  {LeBoeuf}}, \bibinfo {author} {\bibfnamefont {S.}~\bibnamefont {Badoux}},
  \bibinfo {author} {\bibfnamefont {S.}~\bibnamefont {Dufour-Beaus\'ejour}},
  \bibinfo {author} {\bibfnamefont {D.~A.}\ \bibnamefont {Bonn}}, \bibinfo
  {author} {\bibfnamefont {W.~N.}\ \bibnamefont {Hardy}}, \bibinfo {author}
  {\bibfnamefont {R.}~\bibnamefont {Liang}}, \bibinfo {author} {\bibfnamefont
  {D.}~\bibnamefont {Graf}}, \bibinfo {author} {\bibfnamefont {N.}~\bibnamefont
  {Doiron-Leyraud}}, \ and\ \bibinfo {author} {\bibfnamefont {Louis}\
  \bibnamefont {Taillefer}},\ }\bibfield  {title} {\enquote {\bibinfo {title}
  {{Sensitivity of ${T}_{\mathrm{c}}$ to pressure and magnetic field in the
  cuprate superconductor ${\mathrm{YBa}}_{2}{\mathrm{Cu}}_{3}{\mathrm{O}}_{y}$:
  evidence of charge-order suppression by pressure}},}\ }\href@noop {}
  {\bibfield  {journal} {\bibinfo  {journal} {Phys. Rev. B}\ }\textbf {\bibinfo
  {volume} {98}},\ \bibinfo {pages} {064513} (\bibinfo {year}
  {2018})}\BibitemShut {NoStop}%
\bibitem [{\citenamefont {Deng}\ \emph {et~al.}(2019)\citenamefont {Deng},
  \citenamefont {Zheng}, \citenamefont {Wu}, \citenamefont {Huyan},
  \citenamefont {Wu}, \citenamefont {Nie}, \citenamefont {Cho},\ and\
  \citenamefont {Chu}}]{Deng2019}%
  \BibitemOpen
  \bibfield  {author} {\bibinfo {author} {\bibfnamefont {Liangzi}\ \bibnamefont
  {Deng}}, \bibinfo {author} {\bibfnamefont {Yongping}\ \bibnamefont {Zheng}},
  \bibinfo {author} {\bibfnamefont {Zheng}\ \bibnamefont {Wu}}, \bibinfo
  {author} {\bibfnamefont {Shuyuan}\ \bibnamefont {Huyan}}, \bibinfo {author}
  {\bibfnamefont {Hung-Cheng}\ \bibnamefont {Wu}}, \bibinfo {author}
  {\bibfnamefont {Yifan}\ \bibnamefont {Nie}}, \bibinfo {author} {\bibfnamefont
  {Kyeongjae}\ \bibnamefont {Cho}}, \ and\ \bibinfo {author} {\bibfnamefont
  {Ching-Wu}\ \bibnamefont {Chu}},\ }\bibfield  {title} {\enquote {\bibinfo
  {title} {Higher superconducting transition temperature by breaking the
  universal pressure relation},}\ }\href@noop {} {\bibfield  {journal}
  {\bibinfo  {journal} {Proc. Natl. Acad. Sci. U. S. A.}\ }\textbf {\bibinfo
  {volume} {116}},\ \bibinfo {pages} {2004--2008} (\bibinfo {year}
  {2019})}\BibitemShut {NoStop}%
\bibitem [{\citenamefont {Weber}\ \emph {et~al.}(2012)\citenamefont {Weber},
  \citenamefont {Yee}, \citenamefont {Haule},\ and\ \citenamefont
  {Kotliar}}]{Weber2012}%
  \BibitemOpen
  \bibfield  {author} {\bibinfo {author} {\bibfnamefont {C.}~\bibnamefont
  {Weber}}, \bibinfo {author} {\bibfnamefont {C.}~\bibnamefont {Yee}}, \bibinfo
  {author} {\bibfnamefont {K.}~\bibnamefont {Haule}}, \ and\ \bibinfo {author}
  {\bibfnamefont {G.}~\bibnamefont {Kotliar}},\ }\bibfield  {title} {\enquote
  {\bibinfo {title} {{Scaling of the transition temperature of hole-doped
  cuprate superconductors with the charge-transfer energy}},}\ }\href@noop {}
  {\bibfield  {journal} {\bibinfo  {journal} {Europhysics Letters}\ }\textbf
  {\bibinfo {volume} {100}},\ \bibinfo {pages} {37001} (\bibinfo {year}
  {2012})}\BibitemShut {NoStop}%
\bibitem [{\citenamefont {Ruan}\ \emph {et~al.}(2016)\citenamefont {Ruan},
  \citenamefont {Hu}, \citenamefont {Zhao}, \citenamefont {Cai}, \citenamefont
  {Peng}, \citenamefont {Ye}, \citenamefont {Yu}, \citenamefont {Li},
  \citenamefont {Hao}, \citenamefont {Jin}, \citenamefont {Zhou}, \citenamefont
  {Weng},\ and\ \citenamefont {Wang}}]{Ruan2016}%
  \BibitemOpen
  \bibfield  {author} {\bibinfo {author} {\bibfnamefont {Wei}\ \bibnamefont
  {Ruan}}, \bibinfo {author} {\bibfnamefont {Cheng}\ \bibnamefont {Hu}},
  \bibinfo {author} {\bibfnamefont {Jianfa}\ \bibnamefont {Zhao}}, \bibinfo
  {author} {\bibfnamefont {Peng}\ \bibnamefont {Cai}}, \bibinfo {author}
  {\bibfnamefont {Yingying}\ \bibnamefont {Peng}}, \bibinfo {author}
  {\bibfnamefont {Cun}\ \bibnamefont {Ye}}, \bibinfo {author} {\bibfnamefont
  {Runze}\ \bibnamefont {Yu}}, \bibinfo {author} {\bibfnamefont {Xintong}\
  \bibnamefont {Li}}, \bibinfo {author} {\bibfnamefont {Zhenqi}\ \bibnamefont
  {Hao}}, \bibinfo {author} {\bibfnamefont {Changqing}\ \bibnamefont {Jin}},
  \bibinfo {author} {\bibfnamefont {Xingjiang}\ \bibnamefont {Zhou}}, \bibinfo
  {author} {\bibfnamefont {Zheng~Yu}\ \bibnamefont {Weng}}, \ and\ \bibinfo
  {author} {\bibfnamefont {Yayu}\ \bibnamefont {Wang}},\ }\bibfield  {title}
  {\enquote {\bibinfo {title} {{Relationship between the parent charge transfer
  gap and maximum transition temperature in cuprates}},}\ }\href@noop {}
  {\bibfield  {journal} {\bibinfo  {journal} {Science Bulletin}\ ,\ \bibinfo
  {pages} {1--7}} (\bibinfo {year} {2016})}\BibitemShut {NoStop}%
\bibitem [{\citenamefont {Weber}(2017)}]{Weber2017}%
  \BibitemOpen
  \bibfield  {author} {\bibinfo {author} {\bibfnamefont {Cedric}\ \bibnamefont
  {Weber}},\ }\bibfield  {title} {\enquote {\bibinfo {title} {{What controls
  the critical temperature of high temperature copper oxide superconductors:
  insights from scanneling tunnelling microscopy}},}\ }\href@noop {} {\bibfield
   {journal} {\bibinfo  {journal} {Science Bulletin}\ }\textbf {\bibinfo
  {volume} {62}},\ \bibinfo {pages} {102--104} (\bibinfo {year}
  {2017})}\BibitemShut {NoStop}%
\bibitem [{\citenamefont {Sadewasser}\ \emph {et~al.}(1999)\citenamefont
  {Sadewasser}, \citenamefont {Schilling}, \citenamefont {Wagner},
  \citenamefont {Chmaissem}, \citenamefont {Jorgensen}, \citenamefont {Hinks},\
  and\ \citenamefont {Dabrowski}}]{Sadewasser1999}%
  \BibitemOpen
  \bibfield  {author} {\bibinfo {author} {\bibfnamefont {S}~\bibnamefont
  {Sadewasser}}, \bibinfo {author} {\bibfnamefont {J~S}\ \bibnamefont
  {Schilling}}, \bibinfo {author} {\bibfnamefont {J~L}\ \bibnamefont {Wagner}},
  \bibinfo {author} {\bibfnamefont {O}~\bibnamefont {Chmaissem}}, \bibinfo
  {author} {\bibfnamefont {J~D}\ \bibnamefont {Jorgensen}}, \bibinfo {author}
  {\bibfnamefont {D~G}\ \bibnamefont {Hinks}}, \ and\ \bibinfo {author}
  {\bibfnamefont {B}~\bibnamefont {Dabrowski}},\ }\bibfield  {title} {\enquote
  {\bibinfo {title} {{Relaxation effects in the transition temperature of
  superconducting HgBa2CuO4+$\delta$}},}\ }\href@noop {} {\bibfield  {journal}
  {\bibinfo  {journal} {Phys. Rev. B}\ }\textbf {\bibinfo {volume} {60}},\
  \bibinfo {pages} {9827} (\bibinfo {year} {1999})}\BibitemShut {NoStop}%
\bibitem [{\citenamefont {Murayama}\ \emph {et~al.}(1989)\citenamefont
  {Murayama}, \citenamefont {Mori}, \citenamefont {Yomo}, \citenamefont
  {Takagi}, \citenamefont {Uchida},\ and\ \citenamefont
  {Tokura}}]{Murayama1989}%
  \BibitemOpen
  \bibfield  {author} {\bibinfo {author} {\bibfnamefont {C}~\bibnamefont
  {Murayama}}, \bibinfo {author} {\bibfnamefont {N}~\bibnamefont {Mori}},
  \bibinfo {author} {\bibfnamefont {S}~\bibnamefont {Yomo}}, \bibinfo {author}
  {\bibfnamefont {H}~\bibnamefont {Takagi}}, \bibinfo {author} {\bibfnamefont
  {S}~\bibnamefont {Uchida}}, \ and\ \bibinfo {author} {\bibfnamefont
  {Y}~\bibnamefont {Tokura}},\ }\bibfield  {title} {\enquote {\bibinfo {title}
  {{Anomalous absence of pressure effect on transition-temperature in the
  electron-doped superconductor Nd1.85Ce0.15CuO4-$\delta$}},}\ }\href@noop {}
  {\bibfield  {journal} {\bibinfo  {journal} {Nature}\ }\textbf {\bibinfo
  {volume} {339}},\ \bibinfo {pages} {293--294} (\bibinfo {year}
  {1989})}\BibitemShut {NoStop}%
\bibitem [{\citenamefont {Markert}\ \emph {et~al.}(1990)\citenamefont
  {Markert}, \citenamefont {Beille}, \citenamefont {Neumeier}, \citenamefont
  {Early}, \citenamefont {Seaman}, \citenamefont {Moran},\ and\ \citenamefont
  {Maple}}]{Markert1990}%
  \BibitemOpen
  \bibfield  {author} {\bibinfo {author} {\bibfnamefont {J~T}\ \bibnamefont
  {Markert}}, \bibinfo {author} {\bibfnamefont {J}~\bibnamefont {Beille}},
  \bibinfo {author} {\bibfnamefont {J~J}\ \bibnamefont {Neumeier}}, \bibinfo
  {author} {\bibfnamefont {E~A}\ \bibnamefont {Early}}, \bibinfo {author}
  {\bibfnamefont {C~L}\ \bibnamefont {Seaman}}, \bibinfo {author}
  {\bibfnamefont {T}~\bibnamefont {Moran}}, \ and\ \bibinfo {author}
  {\bibfnamefont {M~B}\ \bibnamefont {Maple}},\ }\bibfield  {title} {\enquote
  {\bibinfo {title} {{Pressure-dependence of $T_\text{c}$ in
  L2-\textit{x}M\textit{x}CuO4-\textit{y} (L = Pr,Nd,Sm,Eu M = Ce,Th) -
  antisymmetric behavior of electron-doped versus hole-doped copper-oxide
  superconductors}},}\ }\href@noop {} {\bibfield  {journal} {\bibinfo
  {journal} {Phys. Rev. Lett.}\ }\textbf {\bibinfo {volume} {64}},\ \bibinfo
  {pages} {80--83} (\bibinfo {year} {1990})}\BibitemShut {NoStop}%
\bibitem [{\citenamefont {Ishiwata}\ \emph {et~al.}(2013)\citenamefont
  {Ishiwata}, \citenamefont {Kotajima}, \citenamefont {Takeshita},
  \citenamefont {Terakura}, \citenamefont {Seki},\ and\ \citenamefont
  {Tokura}}]{Ishiwata2013}%
  \BibitemOpen
  \bibfield  {author} {\bibinfo {author} {\bibfnamefont {Shintaro}\
  \bibnamefont {Ishiwata}}, \bibinfo {author} {\bibfnamefont {Daichi}\
  \bibnamefont {Kotajima}}, \bibinfo {author} {\bibfnamefont {Nao}\
  \bibnamefont {Takeshita}}, \bibinfo {author} {\bibfnamefont {Chieko}\
  \bibnamefont {Terakura}}, \bibinfo {author} {\bibfnamefont {Shinichiro}\
  \bibnamefont {Seki}}, \ and\ \bibinfo {author} {\bibfnamefont {Yoshinori}\
  \bibnamefont {Tokura}},\ }\bibfield  {title} {\enquote {\bibinfo {title}
  {{Optimal $T_\text{c}$ for electron-doped cuprate realized under high
  pressure}},}\ }\href@noop {} {\bibfield  {journal} {\bibinfo  {journal}
  {Journal of the Physical Society of Japan}\ }\textbf {\bibinfo {volume}
  {82}},\ \bibinfo {pages} {063705} (\bibinfo {year} {2013})}\BibitemShut
  {NoStop}%
\bibitem [{\citenamefont {Jurkutat}\ \emph {et~al.}(2019)\citenamefont
  {Jurkutat}, \citenamefont {Erb},\ and\ \citenamefont
  {Haase}}]{Jurkutat2019b}%
  \BibitemOpen
  \bibfield  {author} {\bibinfo {author} {\bibfnamefont {Michael}\ \bibnamefont
  {Jurkutat}}, \bibinfo {author} {\bibfnamefont {Andreas}\ \bibnamefont {Erb}},
  \ and\ \bibinfo {author} {\bibfnamefont {J{\"u}rgen}\ \bibnamefont {Haase}},\
  }\bibfield  {title} {\enquote {\bibinfo {title} {{$T_\text{c}$ and other
  cuprate properties in relation to planar charges as measured by NMR}},}\
  }\href@noop {} {\bibfield  {journal} {\bibinfo  {journal} {Condens. Matter}\
  }\textbf {\bibinfo {volume} {4}},\ \bibinfo {pages} {67} (\bibinfo {year}
  {2019})}\BibitemShut {NoStop}%
\bibitem [{\citenamefont {Erb}\ \emph {et~al.}(1996)\citenamefont {Erb},
  \citenamefont {Walker},\ and\ \citenamefont {Fl{\"u}kiger}}]{Erb1996}%
  \BibitemOpen
  \bibfield  {author} {\bibinfo {author} {\bibfnamefont {A}~\bibnamefont
  {Erb}}, \bibinfo {author} {\bibfnamefont {E}~\bibnamefont {Walker}}, \ and\
  \bibinfo {author} {\bibfnamefont {R}~\bibnamefont {Fl{\"u}kiger}},\
  }\bibfield  {title} {\enquote {\bibinfo {title} {{The use of BaZrO3 crucibles
  in crystal growth of the high-$T_\text{c}$ superconductors Progress in
  crystal growth as well as in sample quality}},}\ }\href@noop {} {\bibfield
  {journal} {\bibinfo  {journal} {Physica C: Superconductivity}\ }\textbf
  {\bibinfo {volume} {258}},\ \bibinfo {pages} {9--20} (\bibinfo {year}
  {1996})}\BibitemShut {NoStop}%
\bibitem [{\citenamefont {Forman}\ \emph {et~al.}(1972)\citenamefont {Forman},
  \citenamefont {Piermarini}, \citenamefont {Barnett},\ and\ \citenamefont
  {Block}}]{Forman1972}%
  \BibitemOpen
  \bibfield  {author} {\bibinfo {author} {\bibfnamefont {R~A}\ \bibnamefont
  {Forman}}, \bibinfo {author} {\bibfnamefont {G~J}\ \bibnamefont
  {Piermarini}}, \bibinfo {author} {\bibfnamefont {J~D}\ \bibnamefont
  {Barnett}}, \ and\ \bibinfo {author} {\bibfnamefont {S}~\bibnamefont
  {Block}},\ }\bibfield  {title} {\enquote {\bibinfo {title} {{Pressure
  measurement made by the utilization of ruby sharp-line luminescence.}}}\
  }\href@noop {} {\bibfield  {journal} {\bibinfo  {journal} {Science}\ }\textbf
  {\bibinfo {volume} {176}},\ \bibinfo {pages} {284--285} (\bibinfo {year}
  {1972})}\BibitemShut {NoStop}%
\bibitem [{\citenamefont {Meier}\ \emph {et~al.}(2014)\citenamefont {Meier},
  \citenamefont {Herzig},\ and\ \citenamefont {Haase}}]{Meier2014}%
  \BibitemOpen
  \bibfield  {author} {\bibinfo {author} {\bibfnamefont {T}~\bibnamefont
  {Meier}}, \bibinfo {author} {\bibfnamefont {T}~\bibnamefont {Herzig}}, \ and\
  \bibinfo {author} {\bibfnamefont {J}~\bibnamefont {Haase}},\ }\bibfield
  {title} {\enquote {\bibinfo {title} {{Moissanite anvil cell design for
  giga-pascal nuclear magnetic resonance.}}}\ }\href@noop {} {\bibfield
  {journal} {\bibinfo  {journal} {Rev. Sci. Instrum.}\ }\textbf {\bibinfo
  {volume} {85}},\ \bibinfo {pages} {43903} (\bibinfo {year}
  {2014})}\BibitemShut {NoStop}%
\bibitem [{\citenamefont {Haase}\ \emph {et~al.}(2004)\citenamefont {Haase},
  \citenamefont {Sushkov}, \citenamefont {Horsch},\ and\ \citenamefont
  {Williams}}]{Haase2004}%
  \BibitemOpen
  \bibfield  {author} {\bibinfo {author} {\bibfnamefont {J}~\bibnamefont
  {Haase}}, \bibinfo {author} {\bibfnamefont {O~P}\ \bibnamefont {Sushkov}},
  \bibinfo {author} {\bibfnamefont {P}~\bibnamefont {Horsch}}, \ and\ \bibinfo
  {author} {\bibfnamefont {G~V~M}\ \bibnamefont {Williams}},\ }\bibfield
  {title} {\enquote {\bibinfo {title} {{Planar Cu and O hole densities in
  high-$T_\text{c}$ cuprates determined with NMR}},}\ }\href@noop {} {\bibfield
   {journal} {\bibinfo  {journal} {Phys. Rev. B}\ }\textbf {\bibinfo {volume}
  {69}},\ \bibinfo {pages} {94504} (\bibinfo {year} {2004})}\BibitemShut
  {NoStop}%
\end{thebibliography}%

\end{document}